\documentclass{aa}

\usepackage{graphicx}
\usepackage{wrapfig}
\usepackage{lscape}
\usepackage{color}

\newcommand{\beq}{\begin{equation}}
\newcommand{\eeq}{\end{equation}}
\newcommand{\beqa}{\begin{eqnarray}}
\newcommand{\eeqa}{\end{eqnarray}}

\begin{document}


\title{Numerical simulations of solar macrospicules}

\author{
K. Murawski\inst{1} \and
Abhishek K. Srivastava\inst{2}
\and T.V. Zaqarashvili\inst{3,4}
}

 \institute{Group of Astrophysics,
             UMCS, ul. Radziszewskiego 10, 20-031 Lublin, Poland\\
              \email{kmur@kft.umcs.lublin.pl}
               \and
            Aryabhatta Research Institute of Observational Sciences (ARIES), Manora Peak, Nainital-263 129, Uttarakhand, India
             \email{aks@aries.res.in}
               \and
            Space Research Institute, Austrian Academy of Sciences, Schmiedlstrasse 6, 8042 Graz, Austria\\
             \email{teimuraz.zaqarashvili@oeaw.ac.at}
               \and
            Abastumani Astrophysical Observatory at Ilia State University, University Str. 2, Tbilisi, Georgia
                  }

\date{Received / Accepted }

\abstract
{
We consider a localized pulse in the component of velocity, parallel to the ambient magnetic field lines,
that is
initially launched
in
the solar chromosphere.
}
{
We aim to generalize the recent numerical model of spicule formation (Murawski \& Zaqarashvili 2010) by implementing
a
VAL-C
model of solar temperature.
}
{
With the use of the code FLASH
we solve two-dimensional ideal magnetohydrodynamic equations numerically
to
simulate the solar macrospicules.
}
{
Our numerical
results reveal that
the pulse
located below the transition region
triggers plasma perturbations, which exhibit many features of
macrospicules. We also present an observational (SDO/AIA 304 \AA ) case study of the
macrospicule that approximately mimics the numerical simulations.
}
{
In the frame of the model we devised,
the solar macrospicules can be triggered by velocity pulses launched from
the chromosphere.
}

\keywords{Magnetohydrodynamics (MHD) -- Instabilities -- Sun: atmosphere}

\titlerunning{Solar macrospicules}

\authorrunning{K. Murawski et al.}

\maketitle

\section{Introduction}
%
Spicules are
thin, cool and dense
structures that are observed in the solar limb
(Beckers \cite{bec68,bec72}, Suematsu \cite{Suematsu1998}, Sterling \cite{Sterling2000},
Zaqarashvili \& Erd{\'e}lyi \cite{Zaqarashvili2009}).
%
Type I spicules are seen to be formed at an altitude of about $2000$ km where they reveal a speed of $~25$ km s$^{-1}$, reach at a maximum level
and then either
disappear
or
fall off
to the photosphere.
A typical lifetime of
spicules is $5-15$ min with an average value of $\sim$ $7$ min (Pasachoff et al. \cite{pas09}).
%
Spicules
consist of
double thread structures
(Tanaka \cite{Tanaka1974}, Dara et al. \cite{Dara1998}, Suematsu et al. \cite{Suematsu2008}) and reveal
%
the bi-directional flow (Tsiropoula et al. \cite{Tsiropoula1994}, Tziotziou et al. \cite{Tziotziou2003,Tziotziou2004},
Pasachoff et al. \cite{pas09}).
Typical electron temperature and electron density in spicules are respectively
$(15-17)\cdot 10^3$ K and 2$\cdot 10^{11}-3.5\cdot 10^{10}$ cm$^{-3}$ at altitudes of
4-10 Mm above the solar surface (Beckers \cite{bec68}).
As a result, spicules are much cooler and denser than ambient coronal plasma.
High resolution observations by
Solar Optical Telescope
onboard
Hinode revealed another type of spicules with many features different
than classical limb spicules and they are referred as type II spicules (De Pontieu et al. \cite{dep07}).
Mean diameter of type I spicules is estimated as 660 $\pm$ 200 km (Pasachoff et al. \cite{pas09}), but
the type II spicules have smaller diameters ($\le$ 200 km) in the Ca II H line (De Pontieu et al. \cite{dep07}).
While observed in the H$_\alpha$ line,
type I spicules can reach up to $4-12$ Mm in height from the solar limb with a mean value of $7200$ $\pm$ $2000$ km
(Pasachoff et al. \cite{pas09}). The type II spicules are significantly shorter.
The main differences between type I and type II spicules are: (a) the lifetime, the
former lasts longer; (b) the evolution, the former usually
shows a parabolic profile in Ca II H while the latter
shows an upflow and then disappears; (c) the former usually reveal smaller velocities
(e.g., De Pontieu et al. \cite{dep07,dep09}, Rouppe Van der Voort et al. \cite{rouppe2009}). 

Additionally, very long spicules,
called as {\it macrospicules} with typical length of
up to 40 Mm  and with higher temperature are frequently observed mostly near the polar regions
(Cook et al. \cite{cook84}, Pike \& Harrison \cite{pike97}, Ashbourn \& Woods \cite{ash05}).
Georgakilas et al. (\cite{georg99}) have observed the giant solar spicules that reach at a maximum height of about $12 - 15$ Mm with their lifetime of the order
of $5 - 12$ minutes, and ejection velocity around $\sim$$50$ km s$^{-1}$.
Wilhelm (\cite{klaus2000}) has also observed the largest macrospicules
upto a height of 40 Mm in the north polar coronal hole.
In conclusions, the solar macrospicules can extend from
7 Mm to 45 Mm above the solar limb with lifetime of
3-45 min, and they are mostly concentrated in the polar coronal holes (Sterling \cite{Sterling2000}).

In spite of various theoretical models to explain the spicule ejection in the lower solar atmosphere,
many recent numerical methods have been developed to simulate the
solar spicules/macrospicules with an energy input
at their base in the photosphere as
a pressure pulse or an Alfv\'en wave that steepens into a shock
wave (Sterling \cite{Sterling2000} and references therein). 
Recently, Hansteen et al. (\cite{hansten2006}) and De Pontieu et al. (\cite{dep07}) 
simulated the formation of dynamic fibrils due to slow magneto-acoustic shocks through two-dimensional (2D) numerical 
simulations. They suggest that these shocks are formed when acoustic waves generated by convective flows and global p-modes 
in the lower lying photosphere leak upward into the magnetized chromosphere. Heggland et al. (\cite{heg07}) used 
the initial periodic piston to drive the upward propagating shocks in 1D simulation and 
Martinez-Sykora et al. (\cite{martinez09}) considered the emergence of new magnetic flux 
but the drivers of spicules come from collapsing granules, energy
release in the photosphere or lower chromosphere. 
However, these simulations could not mimic the double structures and
bi-directional flows in spicules. On the other hand, Murawski \& Zaqarashvili (\cite{mur10}) have performed 2D
numerical simulations of magnetohydrodynamic (MHD) equations
and showed that the 2D rebound shock model (Hollweg 1982) may explain both the double structures and bi-directional flows. 
They used a single initial velocity pulse, which leads to the formation of consecutive shocks due to nonlinear wake 
in the stratified atmosphere. However, they considered a simple model of atmospheric temperature 
which was approximated by a smoothed step function profile.

In this paper, we adopt a
more appropriate
temperature profile of Vernazza et al. (\cite{vernaza}) that
implements
characteristic (Alfv\'en and tube) speed profiles and launch the pulse at the chromosphere.
Our more general
model
reveals that macrospicules are effectively excited by velocity pulses launched at the chromosphere.
Additionally, our model confirms the findings of
Murawski \& Zaqarashvili (\cite{mur10}) who
explained the observed properties of spicules (e.g.,
width, height,
and
bi-directional flows). We also present an observational case study using SDO/AIA 304 \AA\
 of a macrospicule in the north polar coronal hole (NPCH), which approximately matches with the simulated macrospicule
in oblique magnetic field.
This  observed spicule appears near the north-east boundary of a moderately evolved hole at the polar cap. 

This paper is organized as follows.
A numerical model is presented in Sect.~2, and the corresponding numerical results
are shown in Sect.~3. The observational case study of a macrospicule is presented in Sect.~4 in support
of the numerical simulations.
This paper
is completed
with a discussion and
conclusions
in Sect.~5.
%
%
%

%
%
\section{A numerical model}\label{sect:num_model}
%
%
%
We consider a gravitationally-stratified solar atmosphere
which is described by
the ideal 2D
MHD equations:
\beqa
\label{eq:MHD_rho}
{{\partial \varrho}\over {\partial t}}+\nabla \cdot (\varrho{\bf V})=0\, ,
\\
\label{eq:MHD_V}
\varrho{{\partial {\bf V}}\over {\partial t}}+ \varrho\left ({\bf V}\cdot \nabla\right ){\bf V} =
-\nabla p+ \frac{1}{\mu}(\nabla\times{\bf B})\times{\bf B} +\varrho{\bf g}\, ,
\\
\label{eq:MHD_p}
{\partial p\over \partial t} + \nabla\cdot (p{\bf V}) = (1-\gamma)p \nabla \cdot {\bf V}\, ,
\hspace{3mm}
p = \frac{k_{\rm B}}{m} \varrho T\, ,
\\
\label{eq:MHD_B}
{{\partial {\bf B}}\over {\partial t}}= \nabla \times ({\bf V}\times{\bf B})\, ,
\hspace{3mm}
\nabla\cdot{\bf B} = 0\, .
\eeqa
Here ${\varrho}$ is mass density, ${\bf V}=[V_{\rm x},V_{\rm y},0]$ is flow velocity,
${\bf B}=[B_{\rm x},B_{\rm y},0]$ is the magnetic field, $p$ is gas pressure, $T$ is temperature,
$\gamma=5/3$ is the adiabatic index, ${\bf g}=(0,-g,0)$ is gravitational acceleration of
its value $g=274$ m s$^{-2}$,
$m$ is mean particle mass and $k_{\rm B}$ is the Boltzmann's constant.
\subsection {Equilibrium configuration}
%
%
%
We assume that the solar atmosphere is in static equilibrium (${\bf V}_{\rm e}={\bf 0}$) with a force-free magnetic field,
\begin{equation}
\label{eq:B}
(\nabla\times{\bf B}_{\rm e})\times{\bf B}_{\rm e} = {\bf 0}\, .
\end{equation}
%
As a result, the pressure gradient is balanced by the gravity force,
\begin{equation}
\label{eq:p}
-\nabla p_{\rm e} + \varrho_{\rm e} {\bf g} = {\bf 0}\, .
\end{equation}
Here the subscript $_{\rm e}$ corresponds to equilibrium quantities.
\begin{figure}[h]
\begin{center}
\includegraphics[scale=0.47, angle=0]{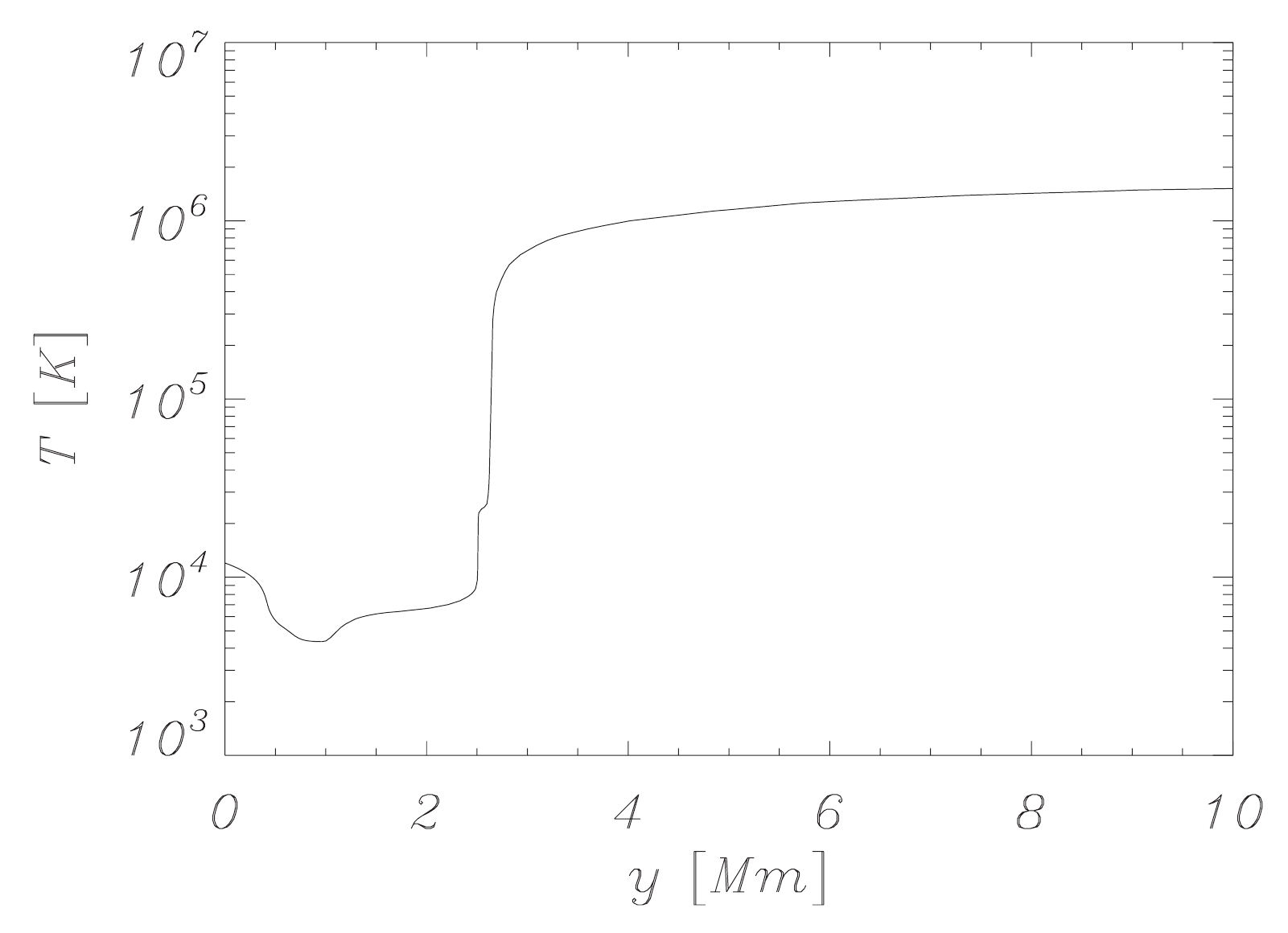}
\includegraphics[scale=0.47, angle=0]{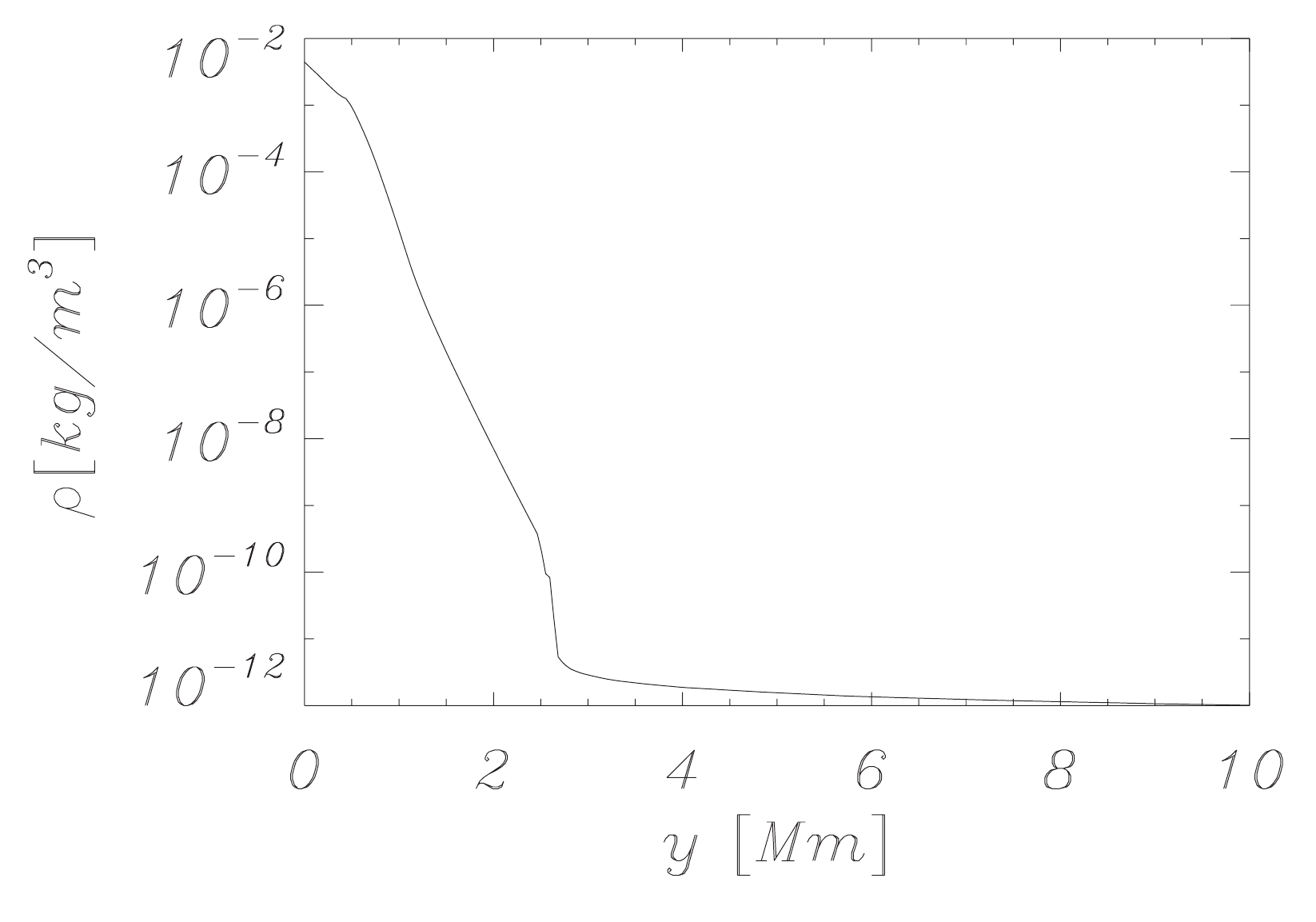}
\caption{\small
Equilibrium profile of solar temperature (top panel) and mass density (bottom panel).
}
\label{fig:initial_profile}
\end{center}
\end{figure}
\begin{figure}[h]
\begin{center}
\includegraphics[width=6.25cm,height=7.25cm, angle=0]{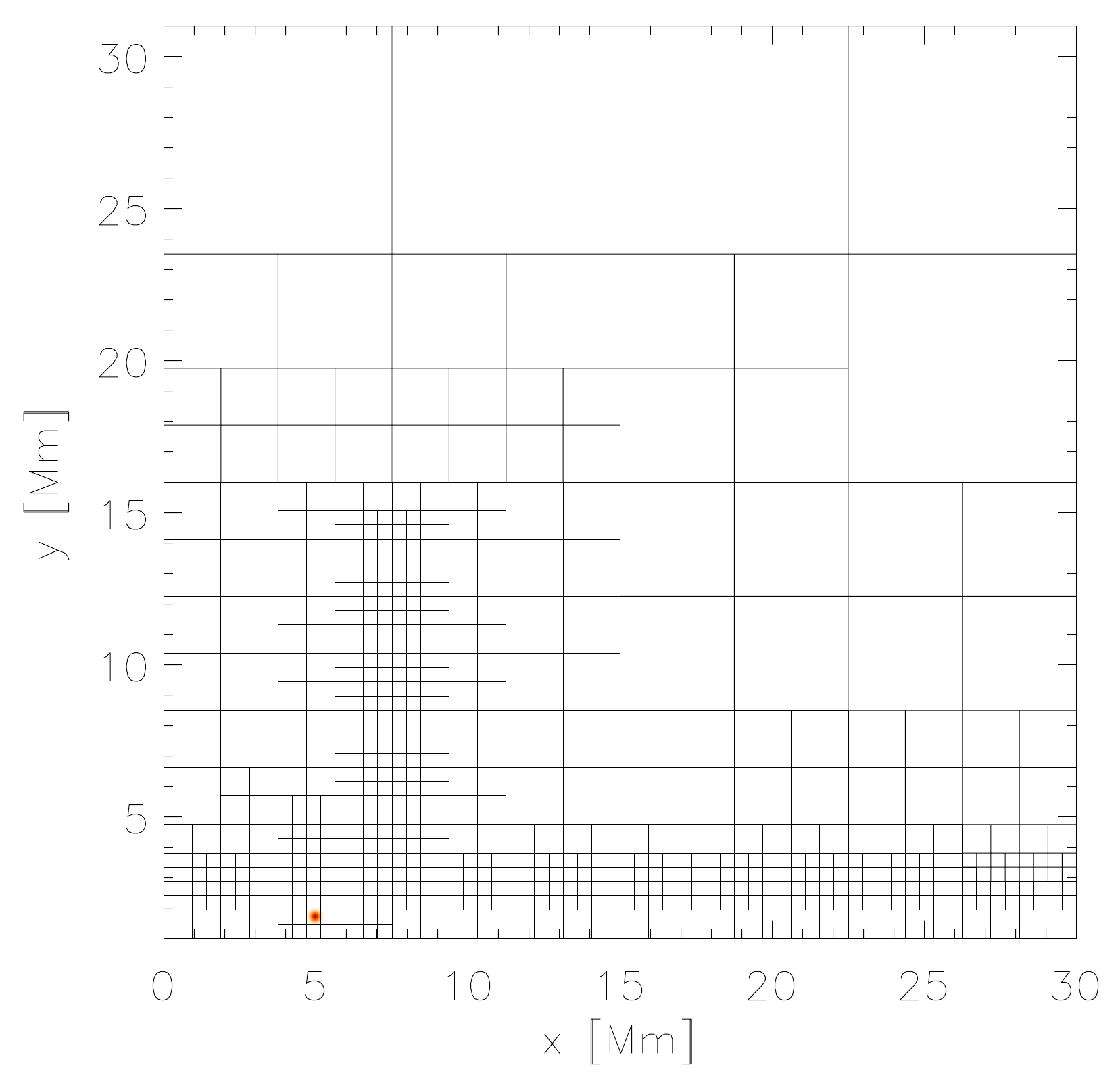}
\caption{\small 
The static grid system with block boundaries represented
by solid lines
for the case of
$x_{\rm 0}=5$ Mm.
The initial perturbation of the system is displayed by 
the red dot at $(5.00,1.75)$ Mm. 
}
\label{fig:amr}
\end{center}
\end{figure}
\begin{figure*}
\centering
\mbox{
\includegraphics[width=5.80cm,height=7.00cm, angle=0]{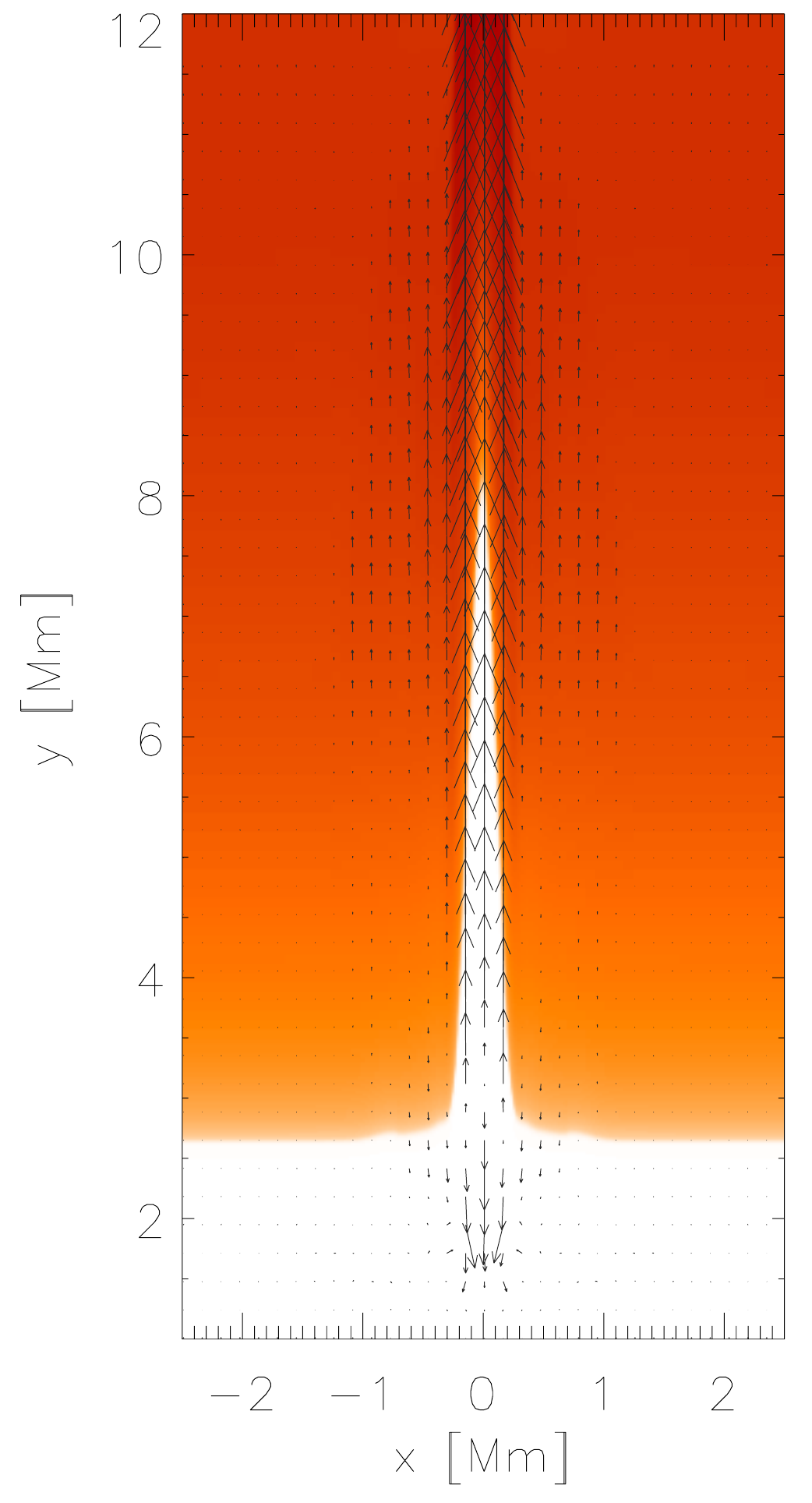}
\includegraphics[width=5.80cm,height=7.00cm, angle=0]{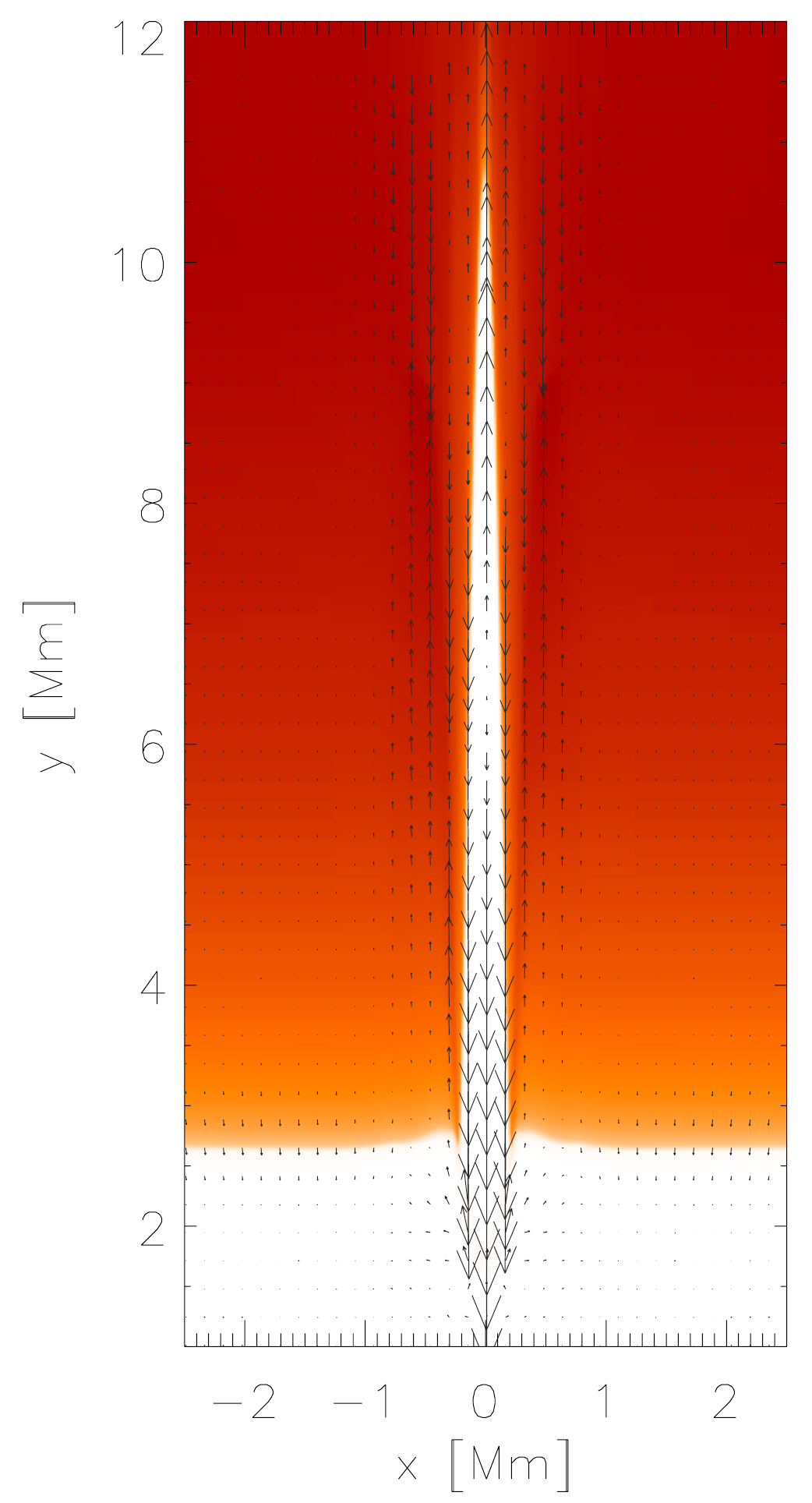}
\includegraphics[width=5.80cm,height=7.00cm, angle=0]{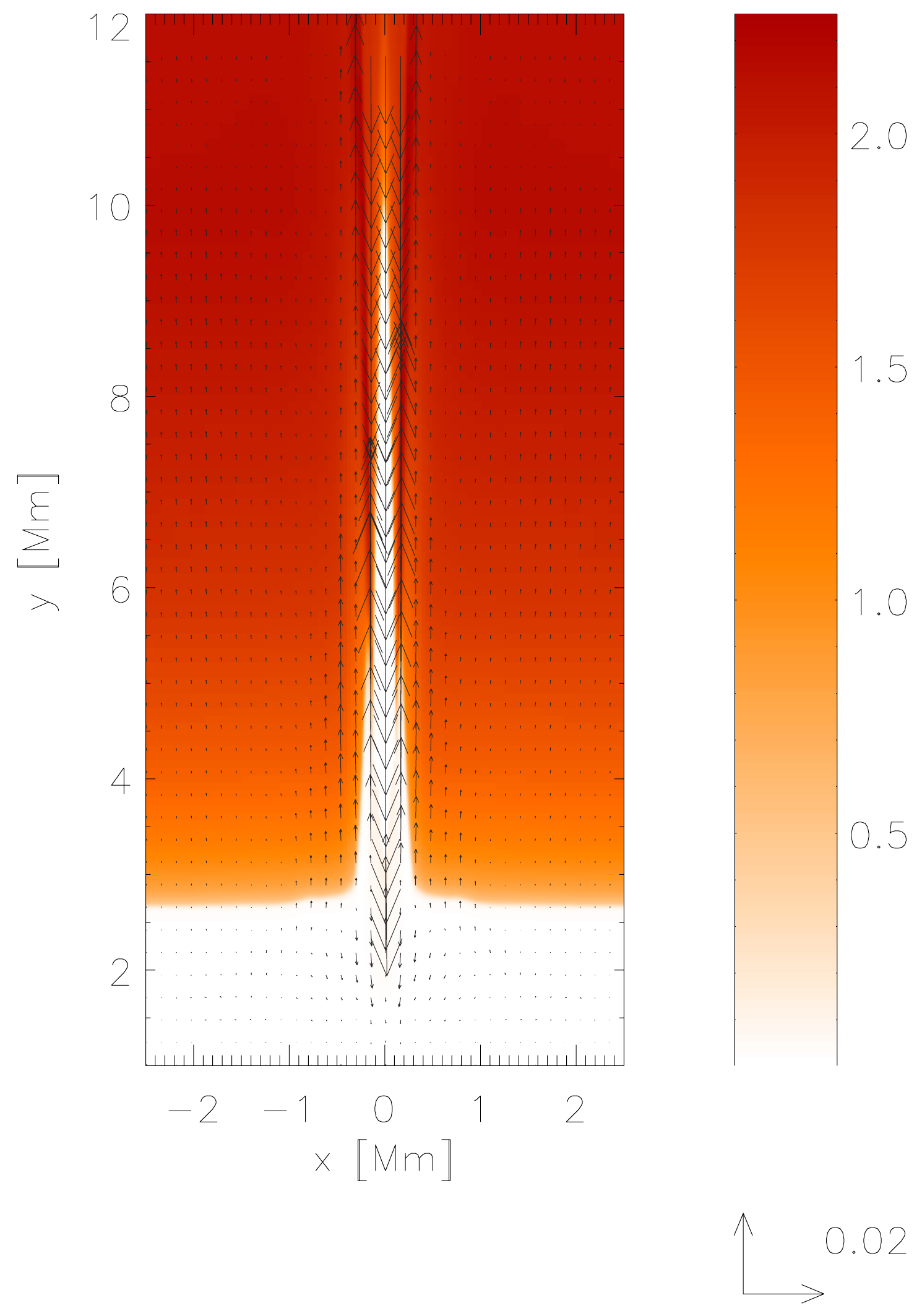}
}
\mbox{
\includegraphics[width=5.80cm,height=7.00cm, angle=0]{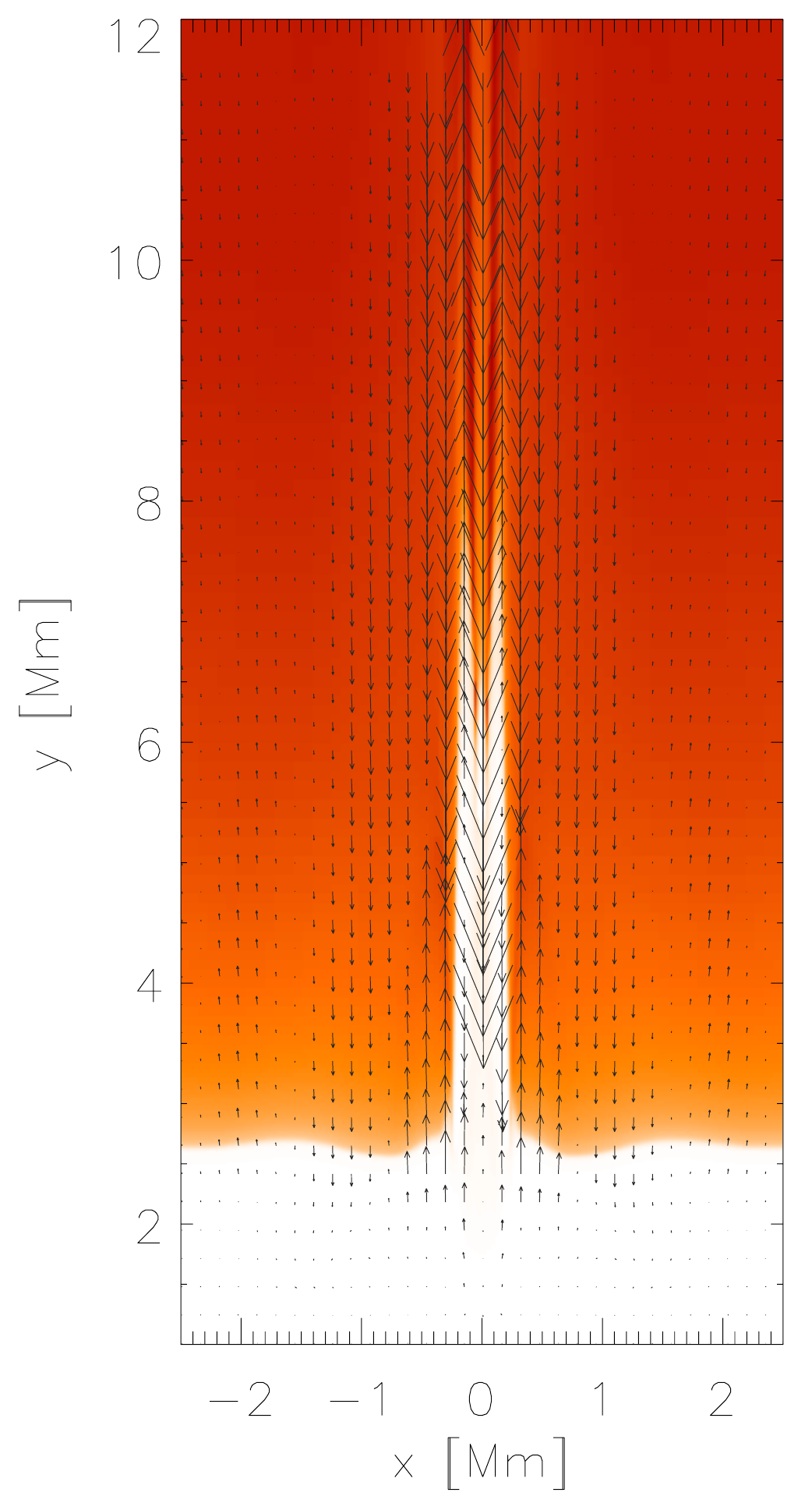}
\includegraphics[width=5.80cm,height=7.00cm, angle=0]{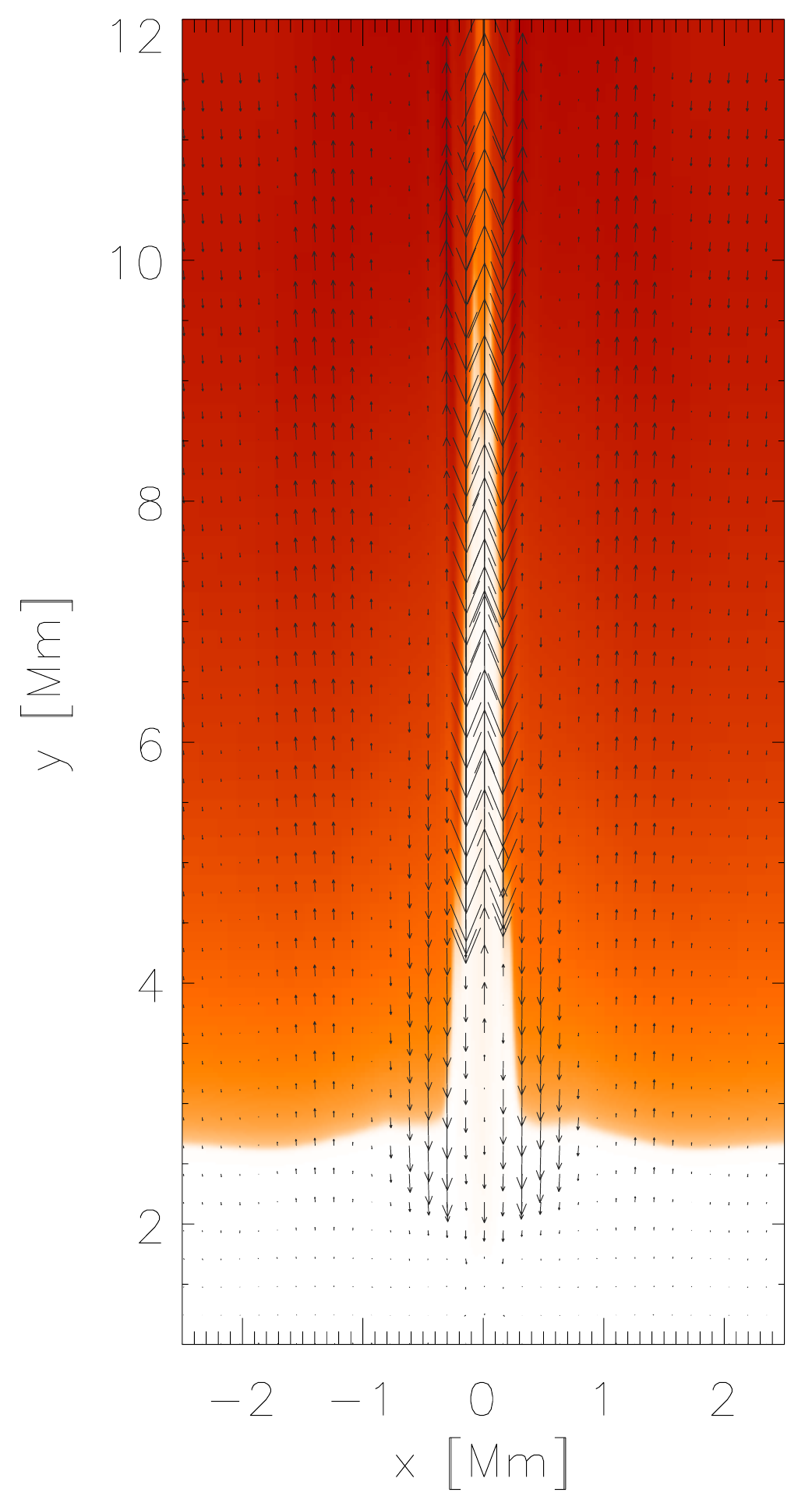}
\includegraphics[width=5.80cm,height=7.00cm, angle=0]{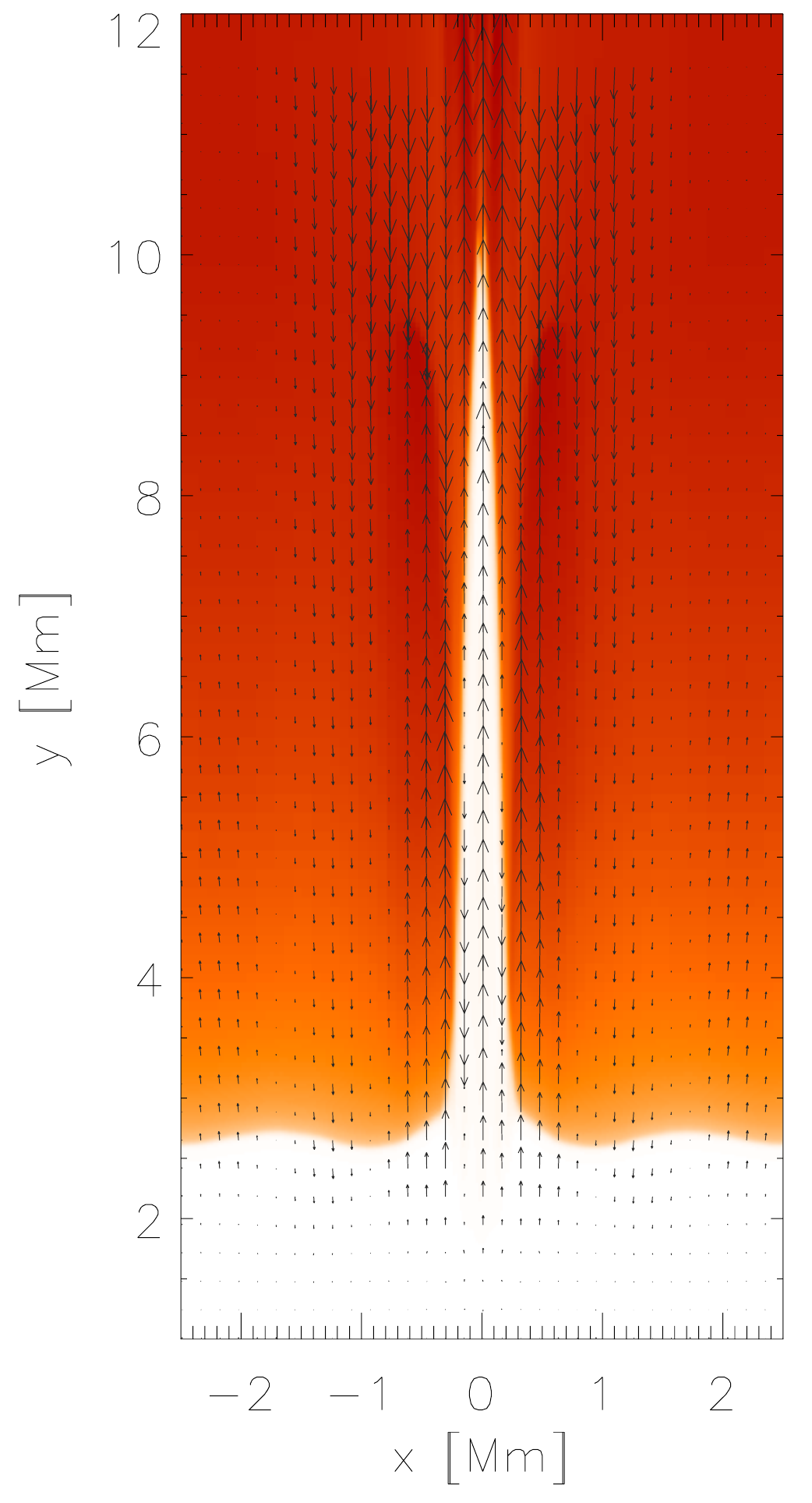}
}
\mbox{
\includegraphics[width=5.80cm,height=7.00cm, angle=0]{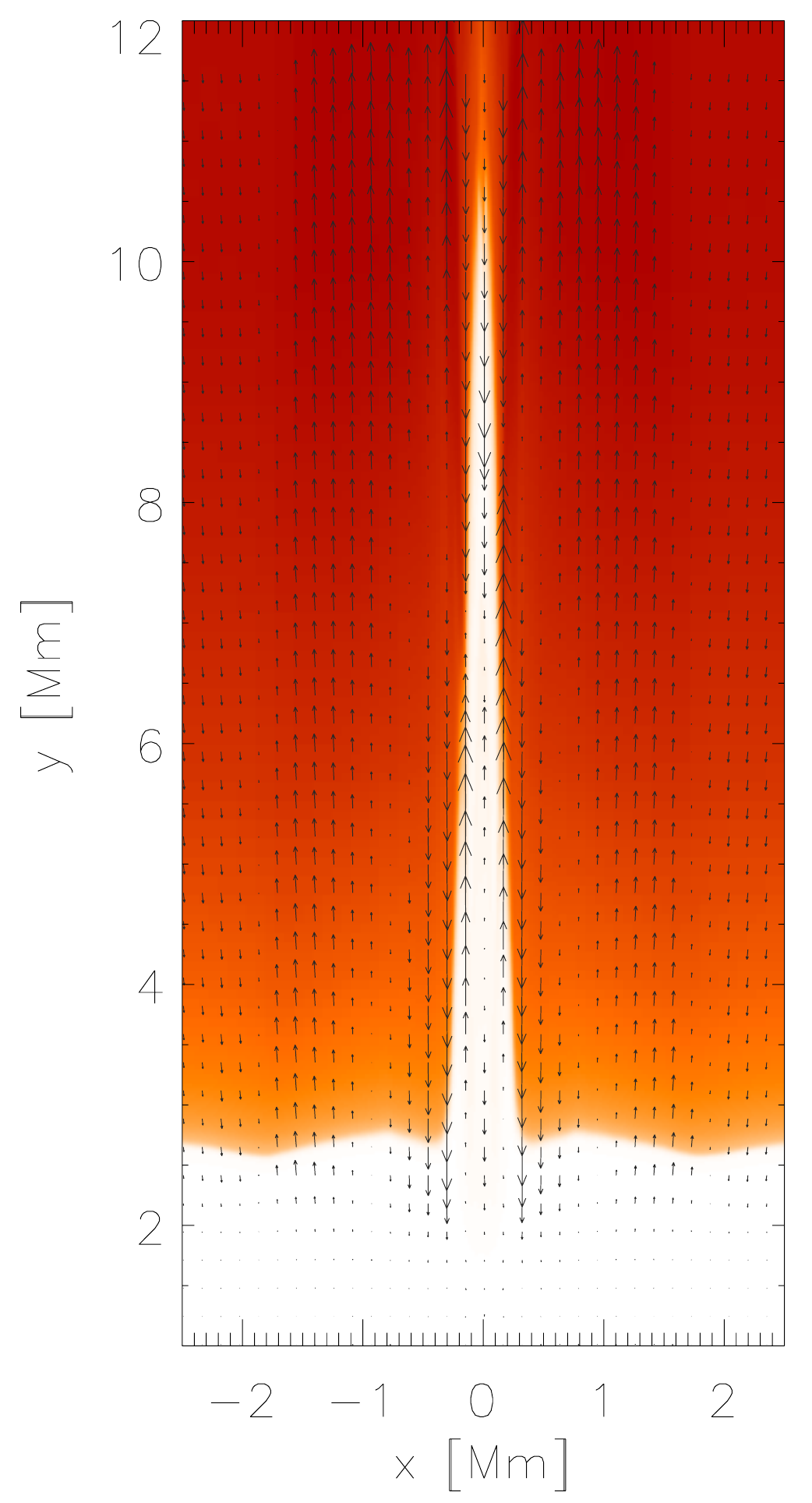}
\includegraphics[width=5.80cm,height=7.00cm, angle=0]{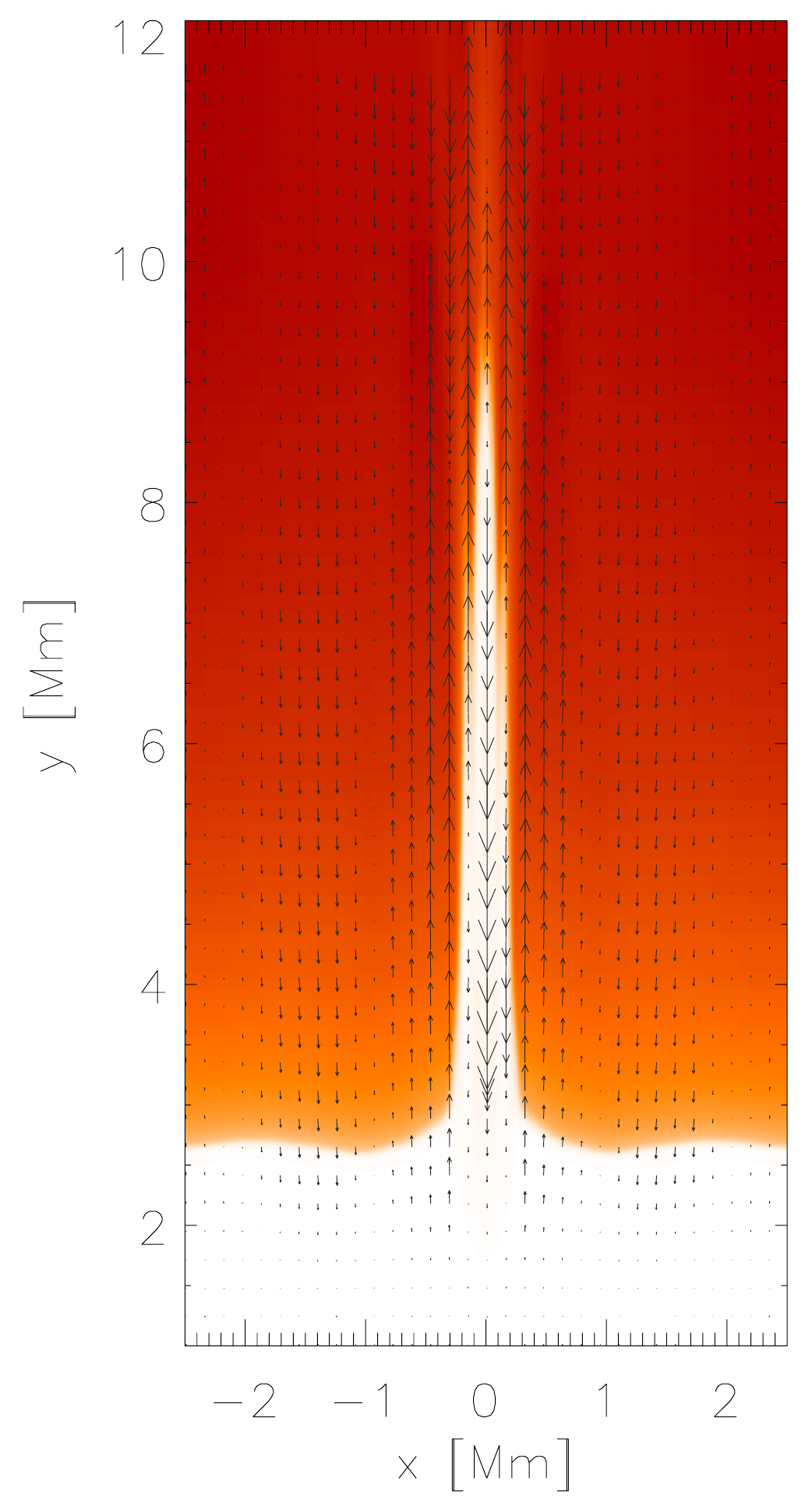}
\includegraphics[width=5.80cm,height=7.00cm, angle=0]{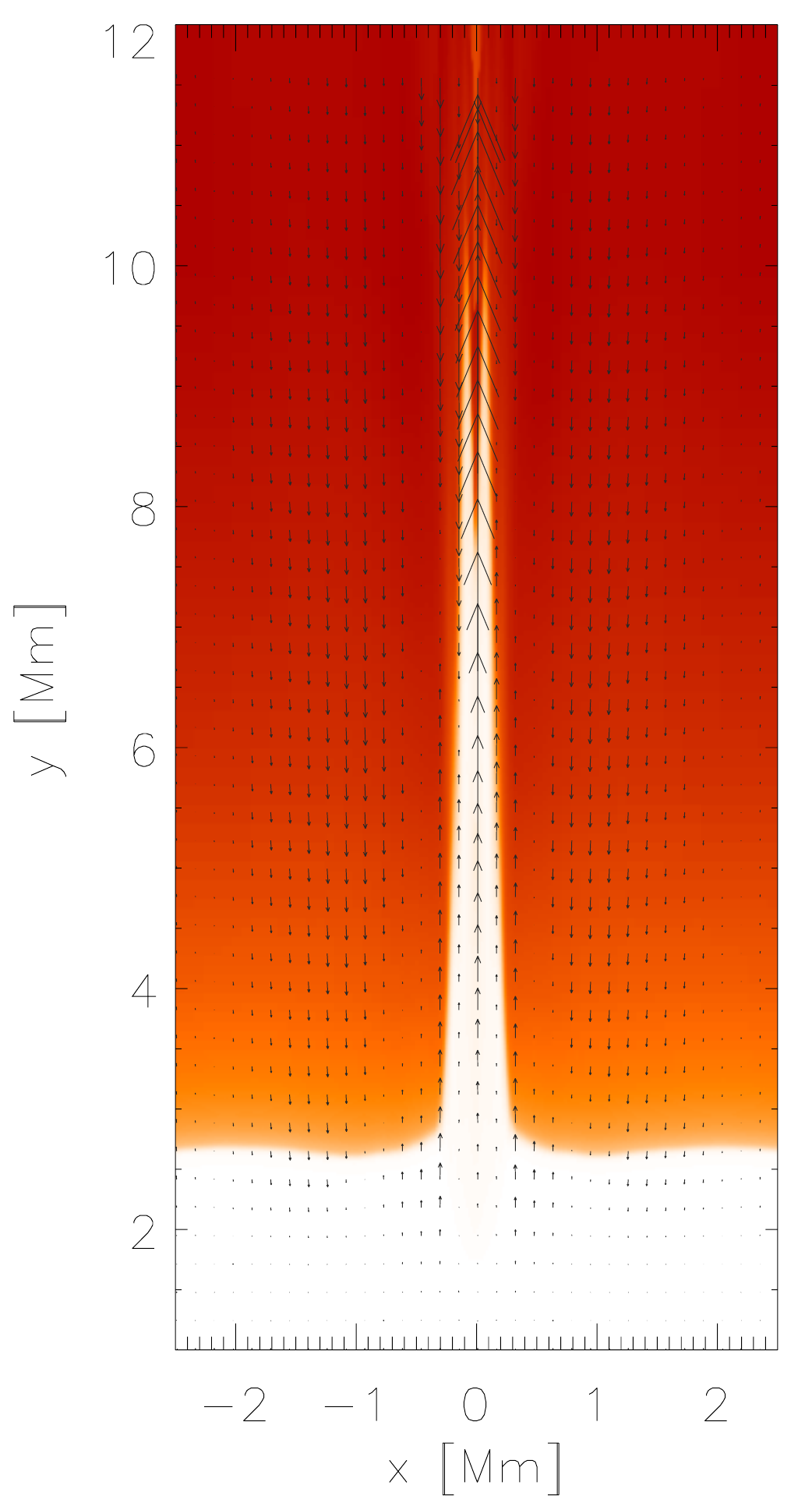}
}
\caption{\small
Temporal snapshots of simulated macrospicule in case of essentially vertical magnetic field (see Sect 3.1). 
The temperature (colour maps) and velocity (arrows) profiles at $t=100$ s,
$t=200$ s, $t=300$ s, $t=400$ s, $t=500$ s, $t=600$ s, $t=700$ s,
$t=800$ s,
and $t=10^3$ s
(from top to bottom) for
$x_{\rm 0}=0$ Mm.
Temperature is drawn in units of $1$ MK.
The arrow below each panel represents the length of the velocity vector, expressed in units of $30$ km s$^{-1}$.
The color bar is common to all the panels.
}
\label{fig:spicule_prof_cent}
\end{figure*}

Using the ideal gas law and the $y$-component of hydrostatic
pressure balance indicated by Eq.~(\ref{eq:p}), we express
equilibrium gas pressure and mass density as
\beqa
\label{eq:pres}
p_{\rm e}(y)=p_{\rm 0}~{\rm exp}\left[ -\int_{y_{\rm r}}^{y}\frac{dy^{'}}{\Lambda (y^{'})} \right]\, ,\hspace{3mm}
\label{eq:eq_rho}
\varrho_{\rm e} (y)=\frac{p_{\rm e}(y)}{g \Lambda(y)}\, .
\eeqa
Here
\begin{equation}
\Lambda(y) = k_{\rm B} T_{\rm e}(y)/(mg)
\end{equation}
is the pressure scale-height, and $p_{\rm 0}$ denotes the gas
pressure at the reference level that we choose in the solar corona at $y_{\rm r}=10$ Mm.

We adopt
an
equilibrium temperature profile $T_{\rm e}(z)$ for the solar atmosphere
that is close to the VAL-C atmospheric model of Vernazza et al. (1981). It is smoothly extended into the corona (Fig.~\ref{fig:initial_profile}, top).
Then with the use of Eq.~(\ref{eq:pres})
we obtain the corresponding gas pressure and mass density profiles.

We assume that the initial magnetic field satisfies a current-free condition,
%
$\nabla \times \vec B_{\rm e}={\bf 0}$, and it is specified by the magnetic flux function, $A$,
such that
\beq
\vec B_{\rm e}(x,y) = \left[\frac{\partial A}{\partial y}, -\frac{\partial A}{\partial x},0\right]
\eeq
%
with
\begin{equation}
A(x,y) = S_{\rm 1} \frac{x-x_{\rm 1}}{(x-x_{\rm 1})^2+(y-y_{\rm 1})^2}\, .
\end{equation}
Here $S_{\rm 1}$ denotes a strength of magnetic moment that is located at ($x_{\rm 1}$, $y_{\rm 1}$).
We choose and hold fixed $x_{\rm 1}=0$ Mm, $y_{\rm 1}=-40$ Mm, and $S_{\rm 1}$ is specified from the requirement that
at the reference point ($x_{\rm r}=0$, $y_{\rm r}=10$) Mm Alfv\'en speed,
$c_{\rm A}(x_{\rm r},y_{\rm r})= {\vert {\bf B_{\rm e}}(x_{\rm r},y_{\rm r})\vert} / {\sqrt{\mu \varrho_{\rm e}(y_{\rm r})}}$,
is $10$ times larger than the sound speed, $c_{\rm s}(y_{\rm r})=\sqrt{{\gamma p_{\rm e}(y_{\rm r})} /  {\varrho_{\rm e}(y_{\rm r})}}$.

\begin{figure}[!!h]
\begin{center}
\includegraphics[scale=0.45]{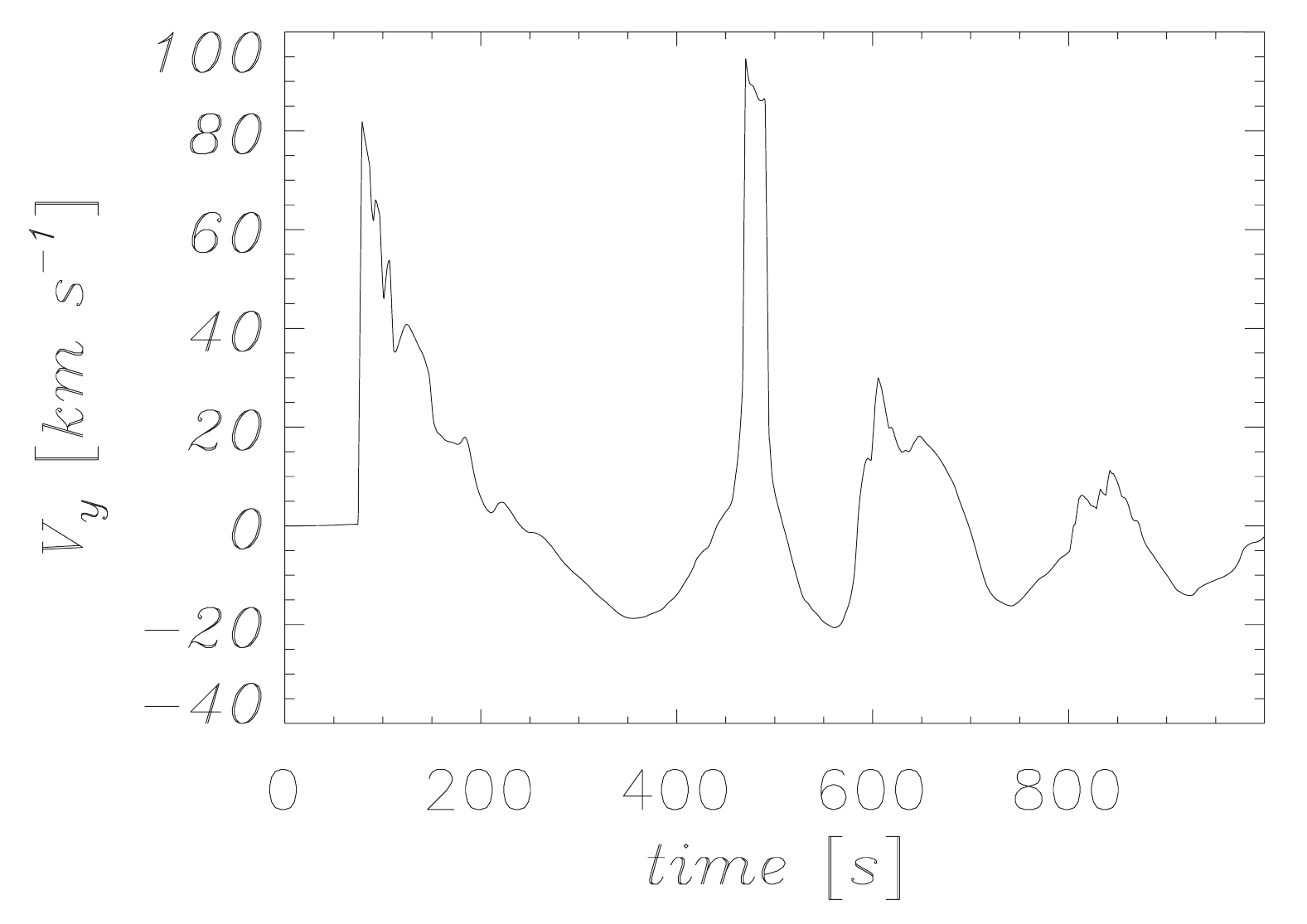}
\caption{\small
Time signature of 
velocity $V_{\rm y}$ collected at ($x=0, y=14)$ Mm 
for the case of essentially vertical magnetic field as shown in Fig.~3.
}
\label{fig:time_profile}
\end{center}
\end{figure}
%

%
%
%
%
%
%
%
For
these settings magnetic field
is predominantly vertical around $x=0$ Mm, while around ($x=5$, $y=1.75$) Mm it is oblique with about $\pi/3$ angle to the solar surface.


%
\begin{figure*}
\centering
\mbox{
\includegraphics[width=5.80cm,height=7.00cm, angle=0]{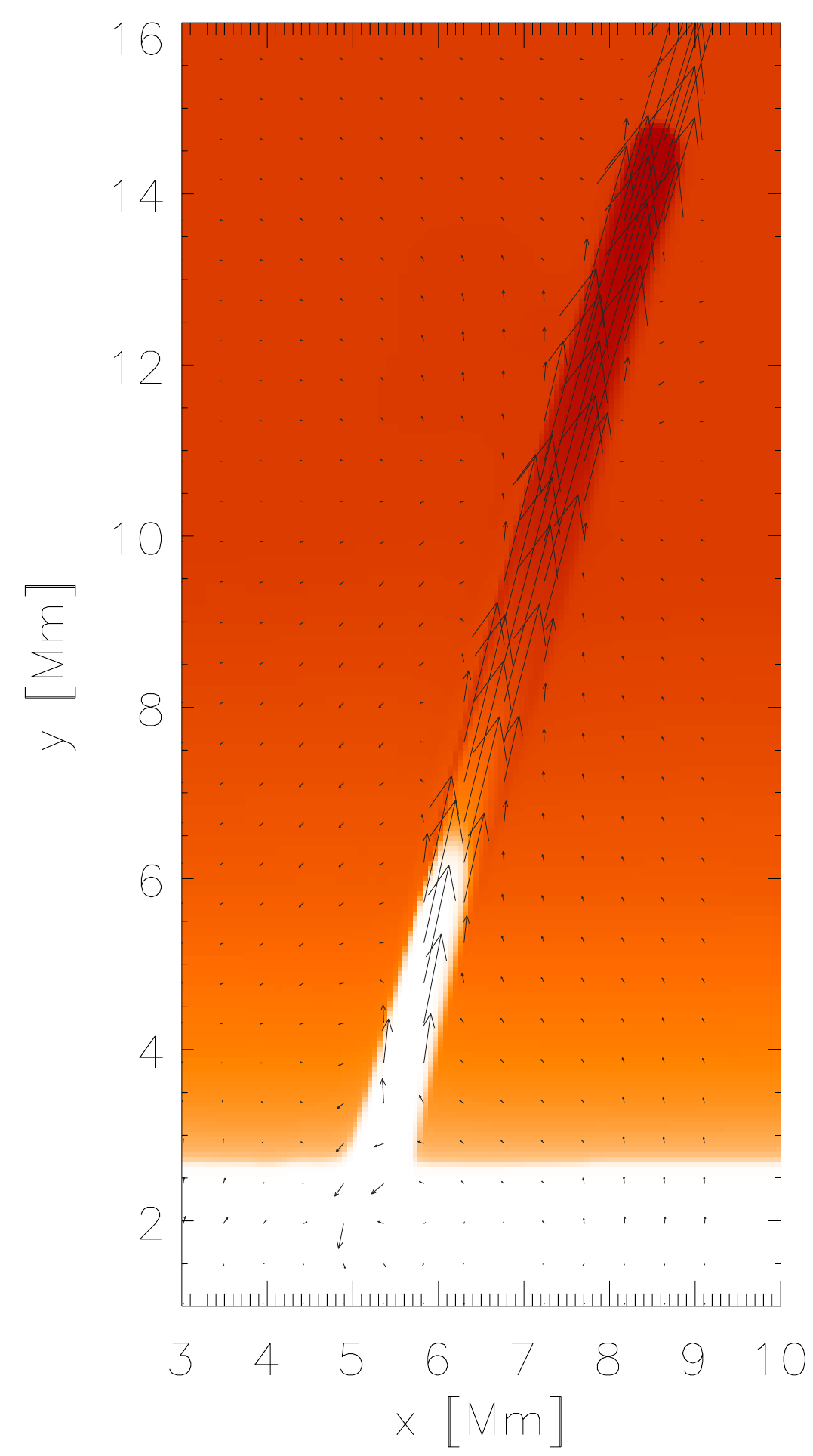}
\includegraphics[width=5.80cm,height=7.00cm, angle=0]{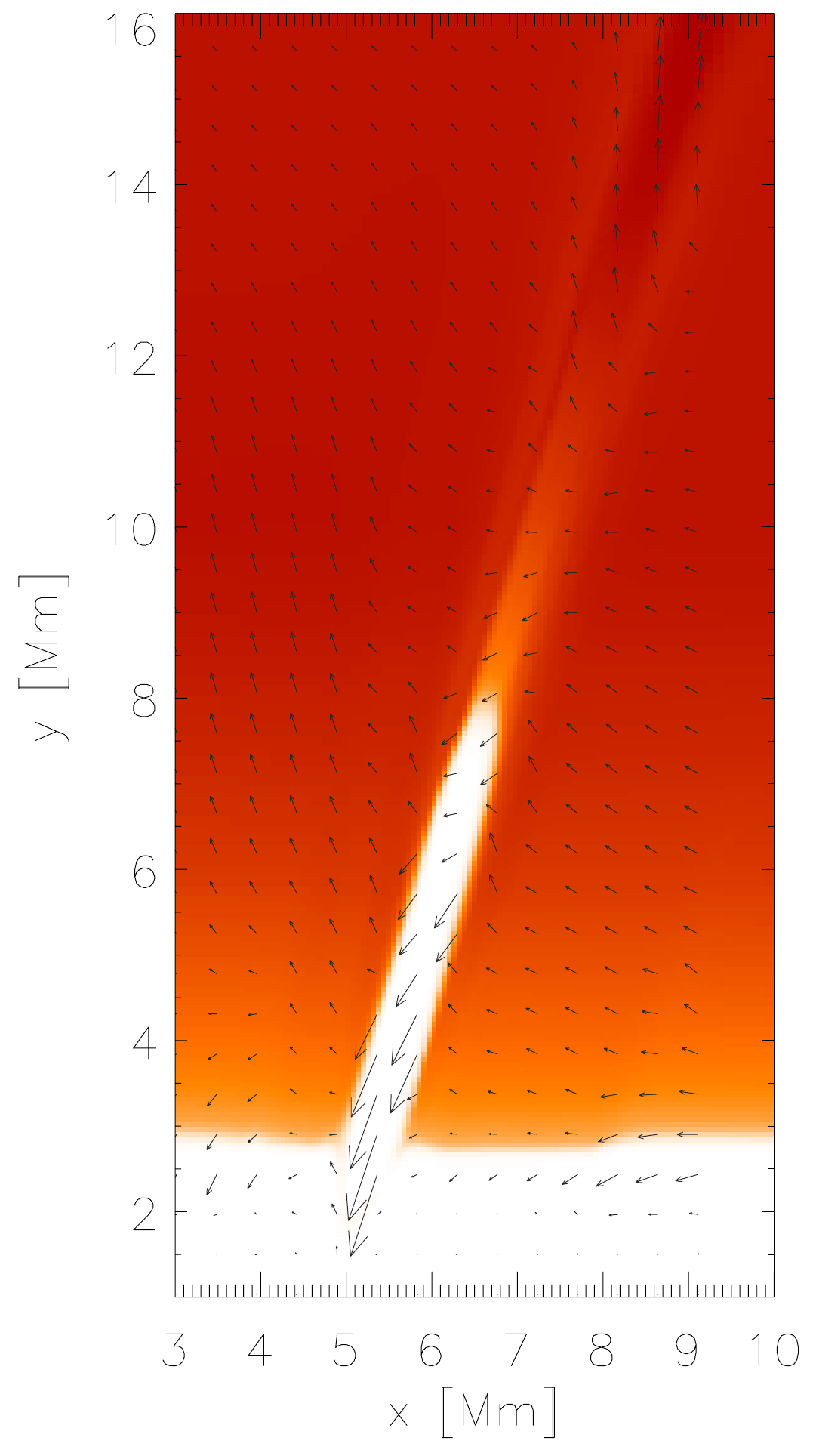}
\includegraphics[width=5.80cm,height=7.00cm, angle=0]{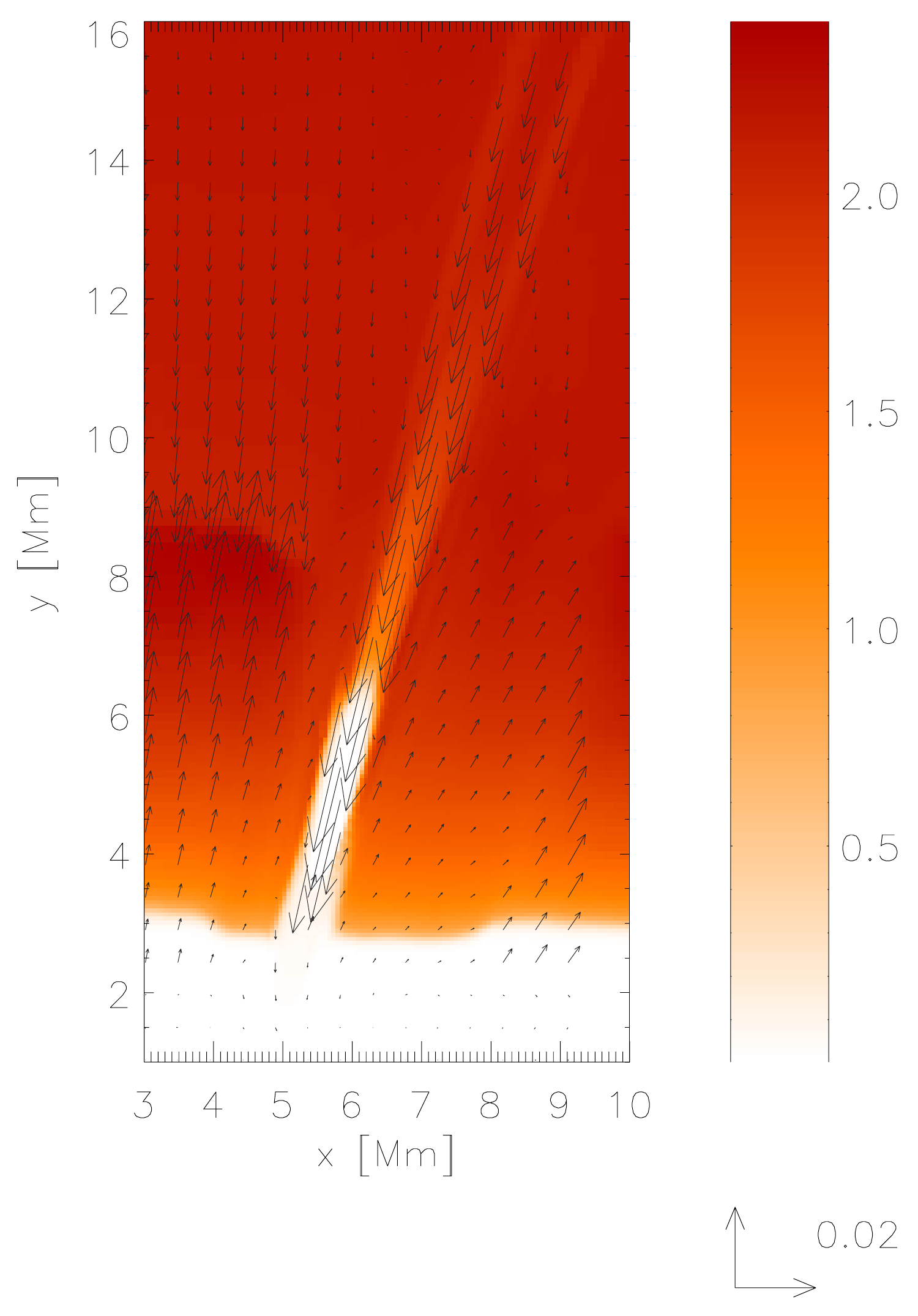}
}
\mbox{
\includegraphics[width=5.80cm,height=7.00cm, angle=0]{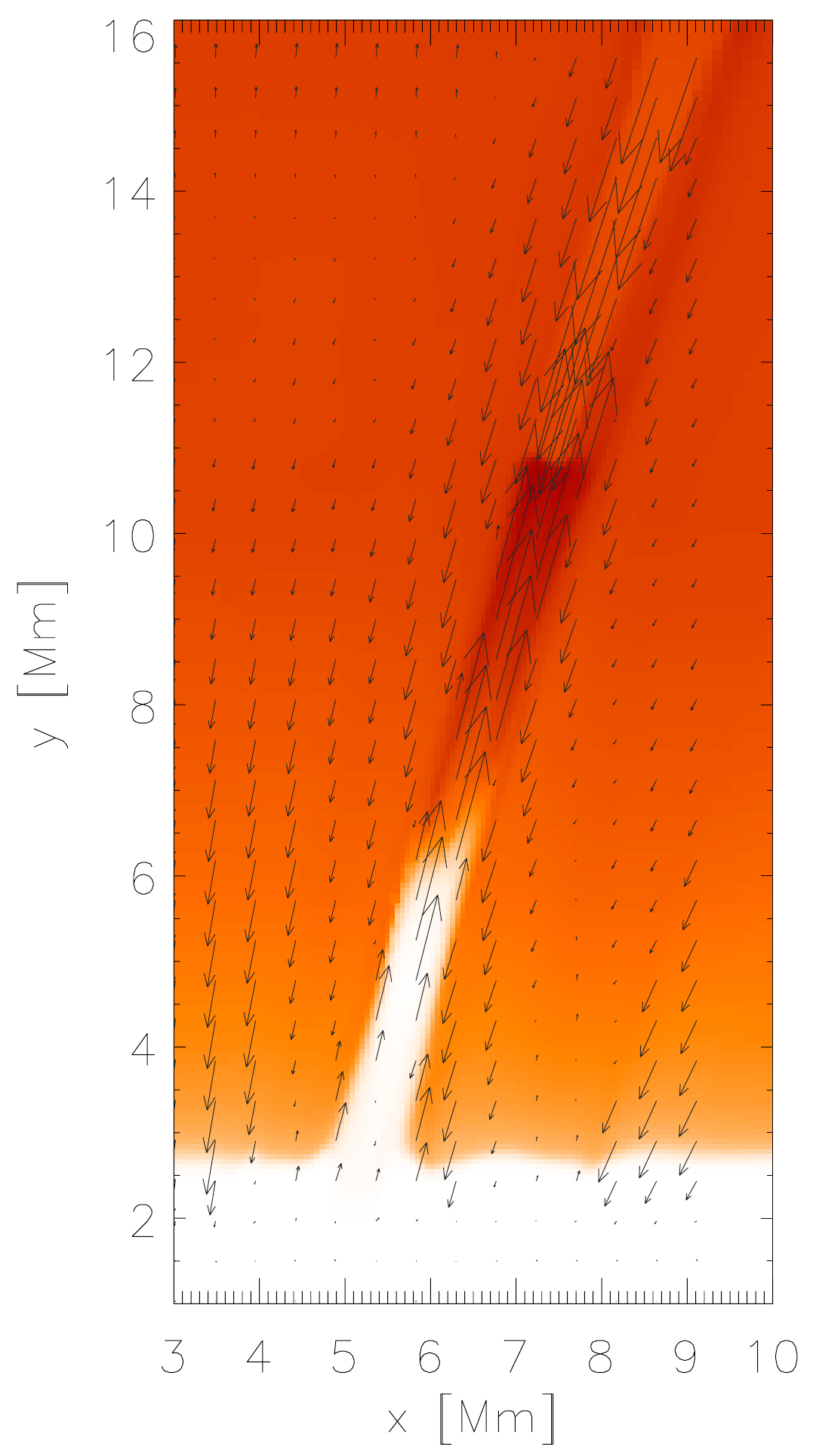}
\includegraphics[width=5.80cm,height=7.00cm, angle=0]{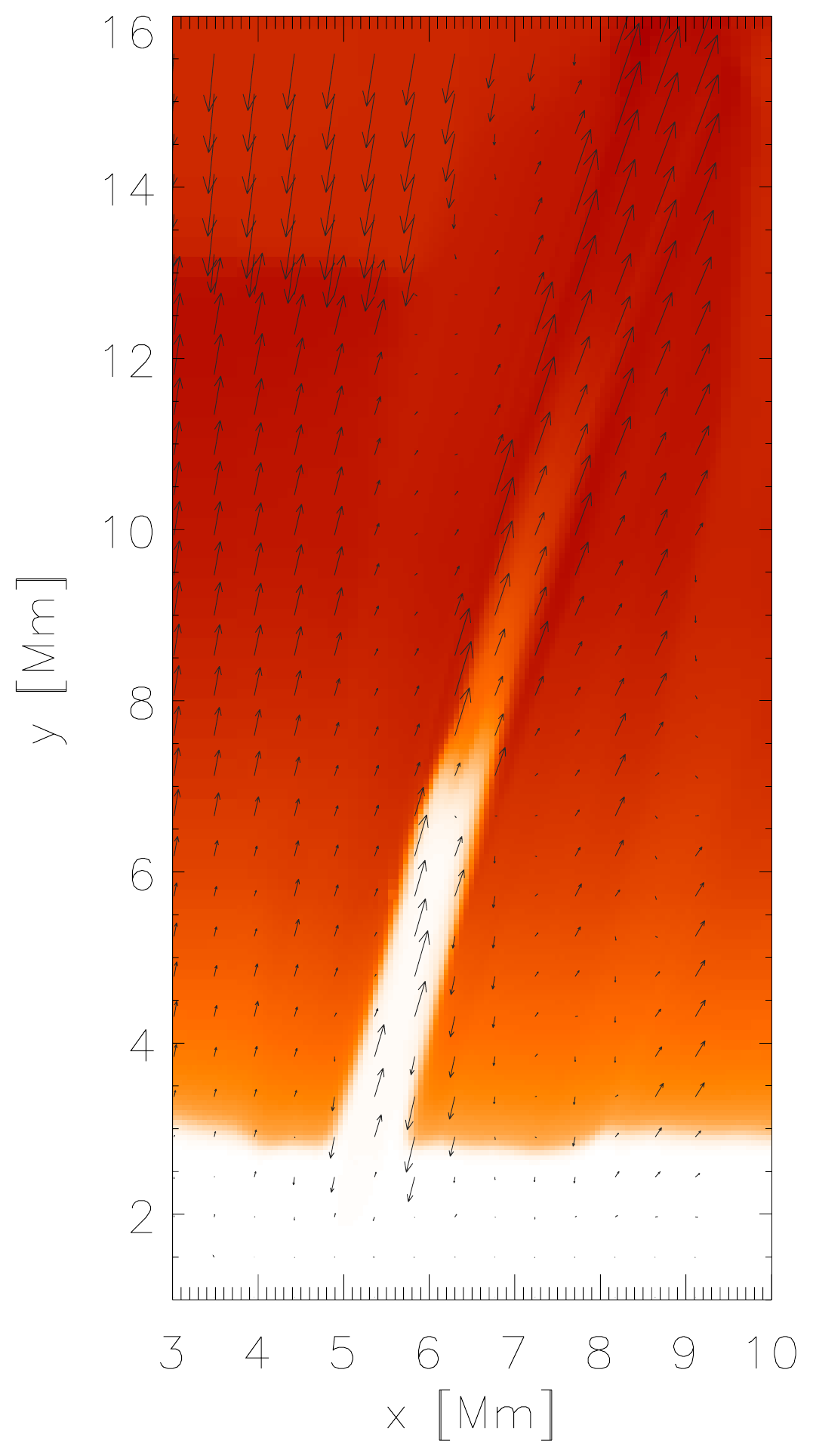}
\includegraphics[width=5.80cm,height=7.00cm, angle=0]{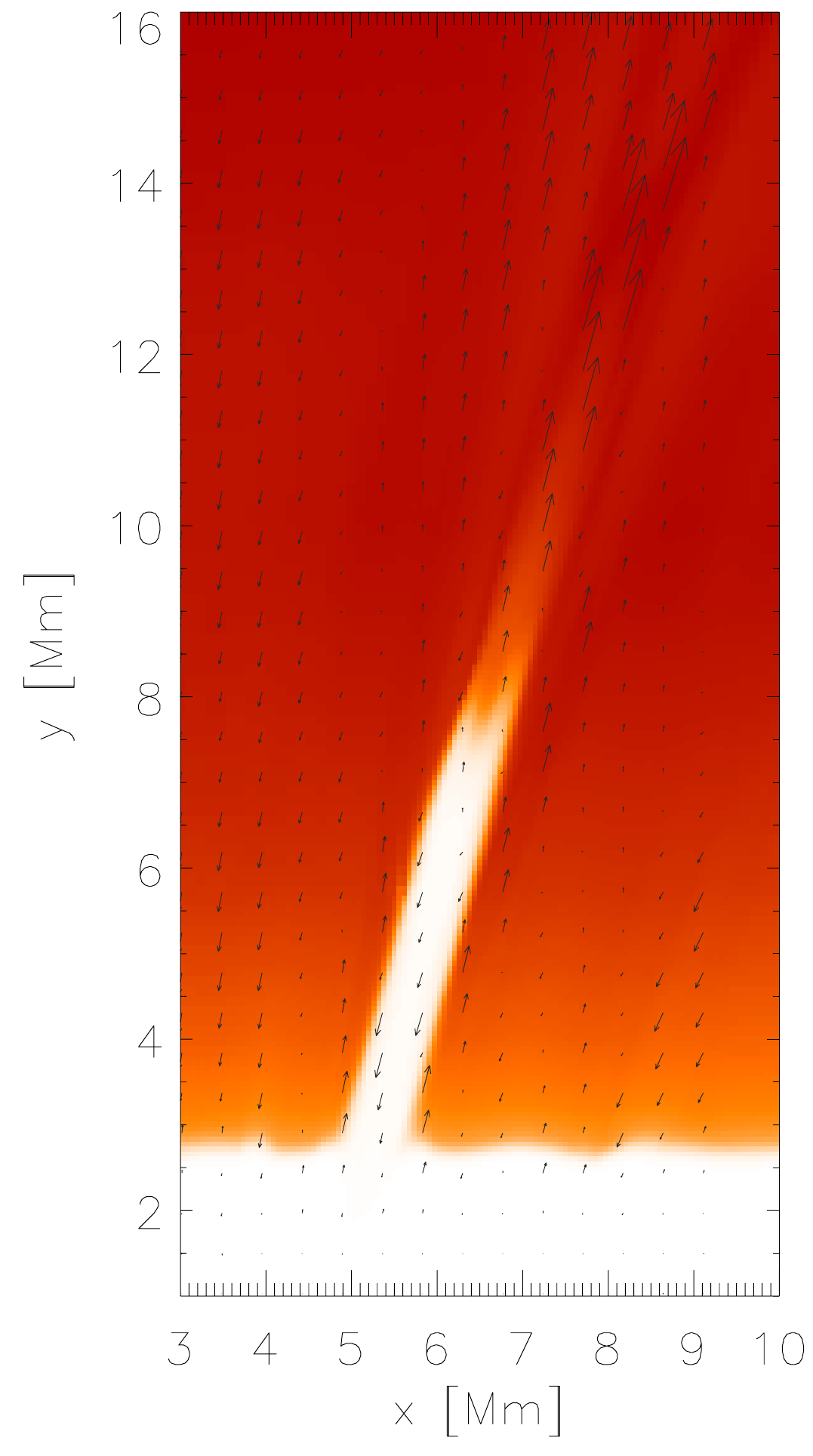}
}
\mbox{
\includegraphics[width=5.80cm,height=7.00cm, angle=0]{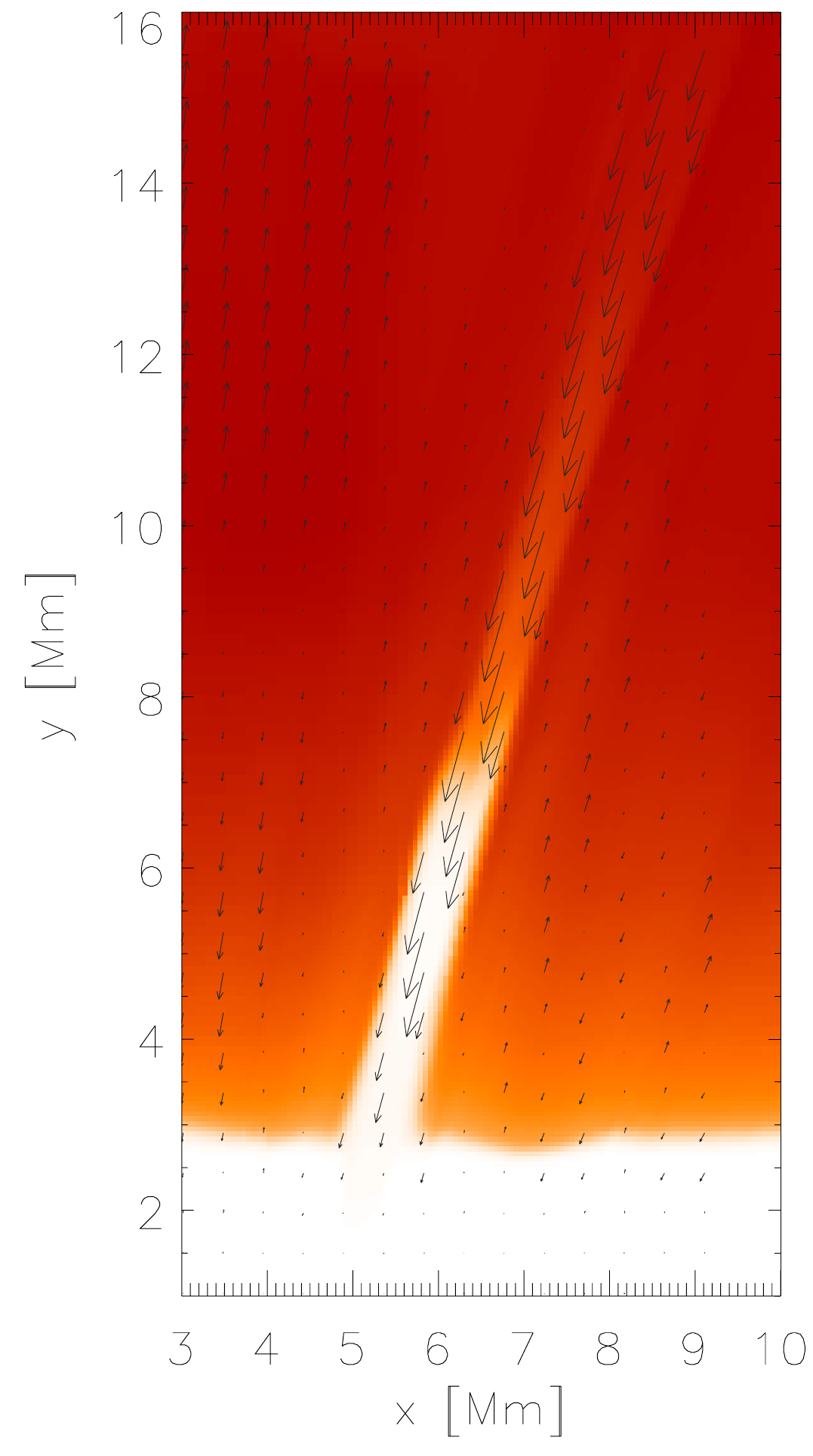}
\includegraphics[width=5.80cm,height=7.00cm, angle=0]{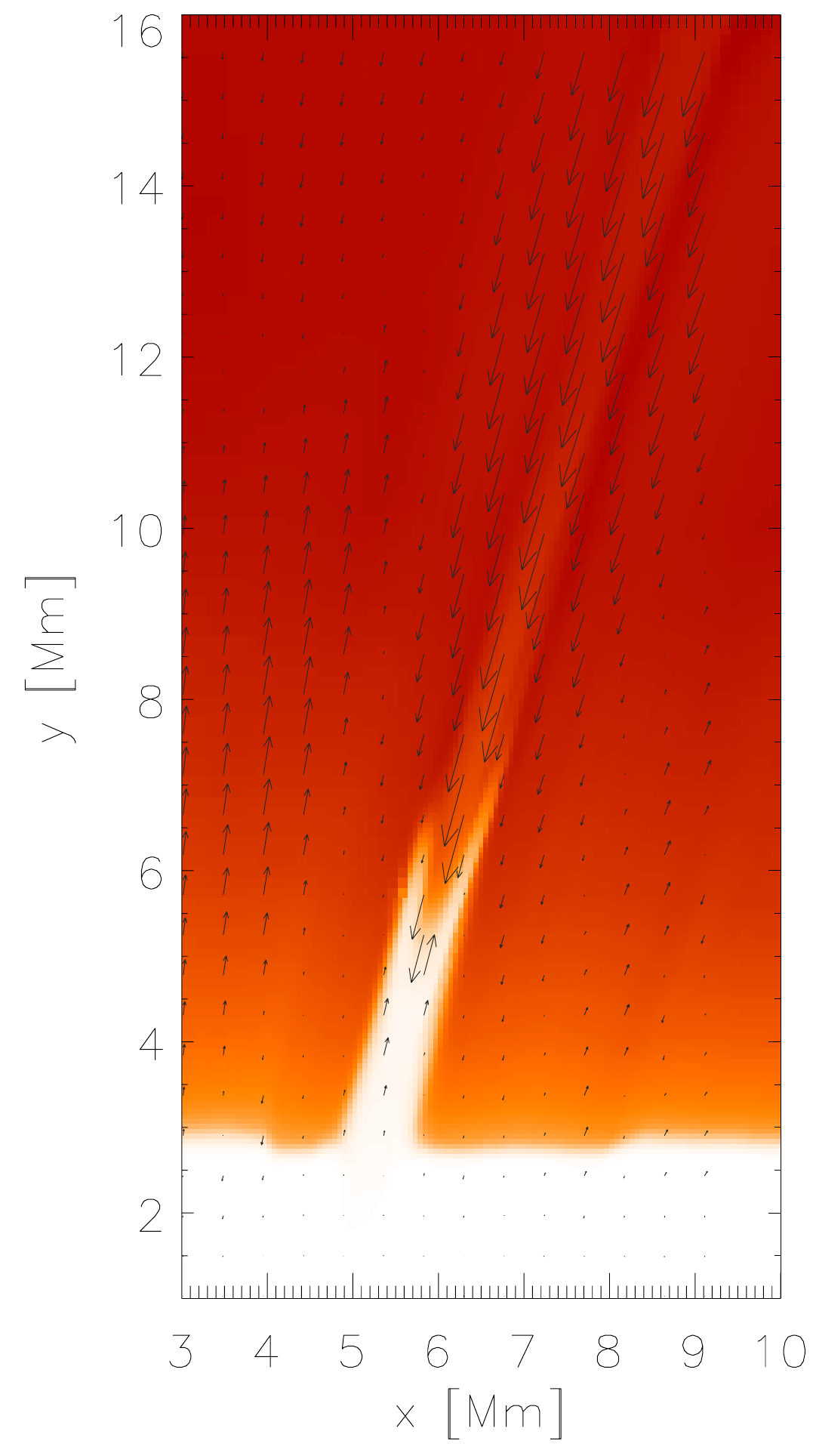}
\includegraphics[width=5.80cm,height=7.00cm, angle=0]{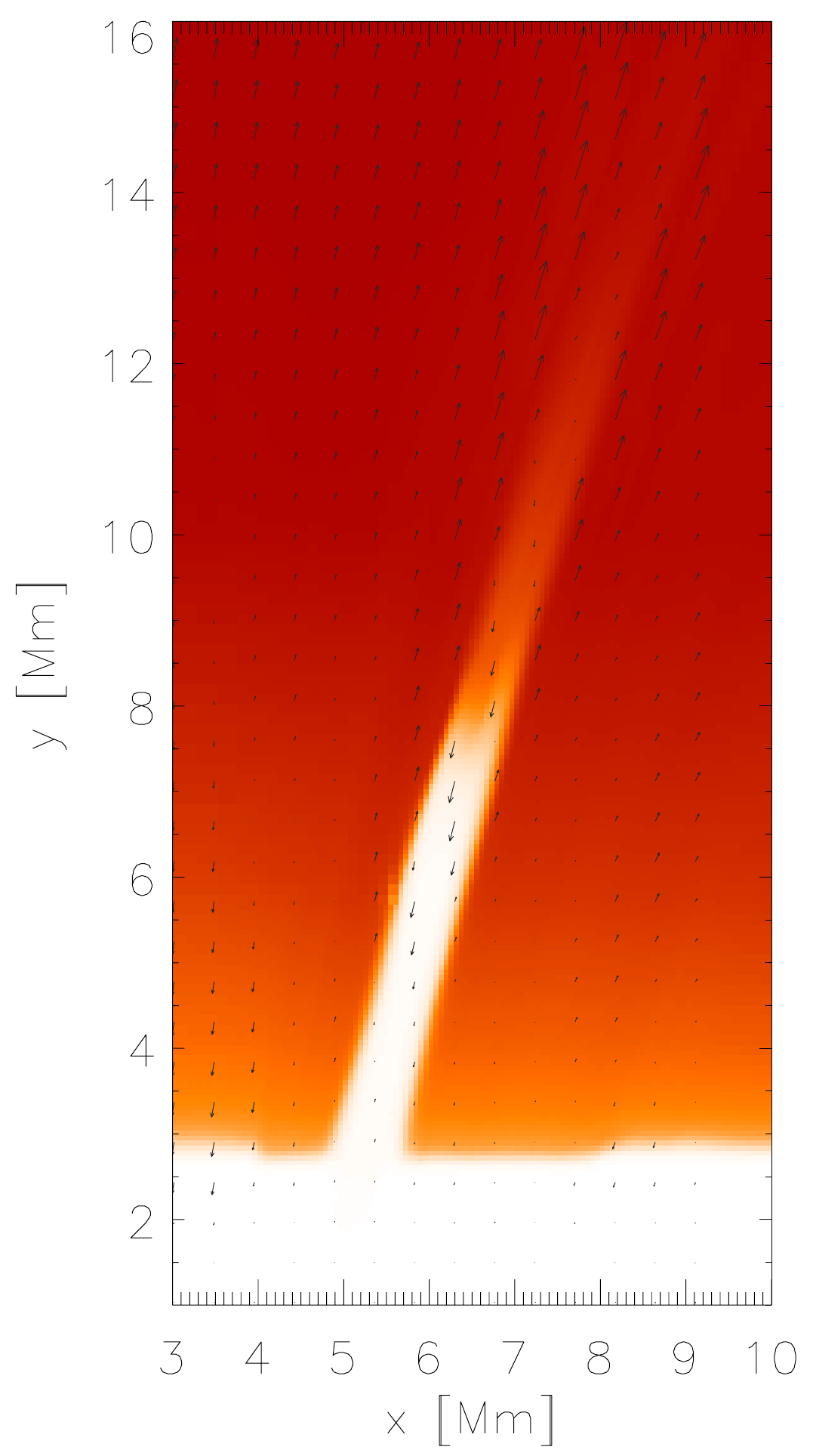}
}
\caption{\small
Same as Fig. 3, but for the case of oblique magnetic field (See Sect.~3.2).
}
\label{fig:spicule_prof_x0=5}
\end{figure*}
\begin{figure}[!h]
\begin{center}
\includegraphics[scale=0.45]{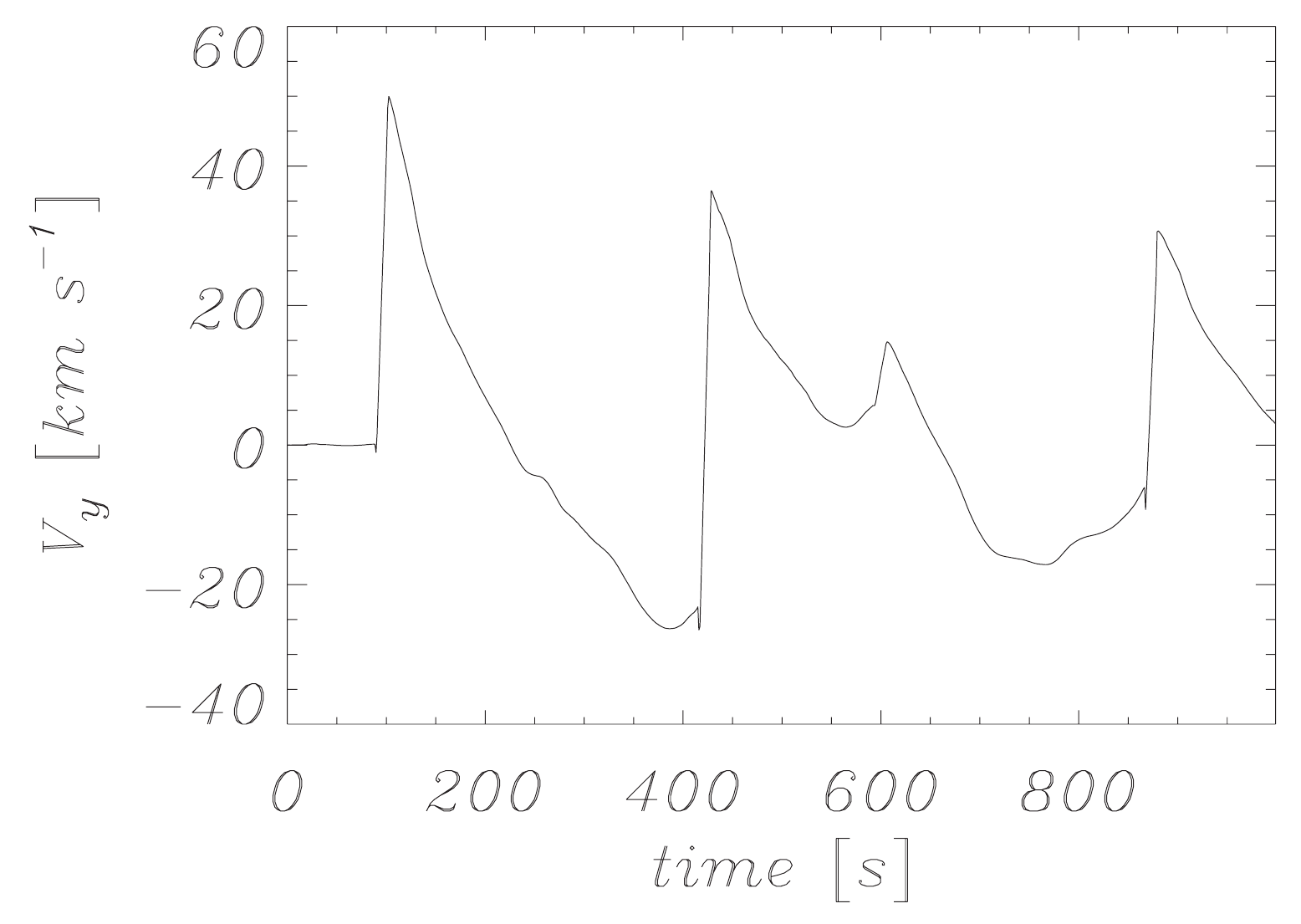}
\caption{\small
Same as Fig.~4, but for the case of oblique magnetic field shown in Fig.~5. 
}
\label{fig:time_profile_x0=5}
\end{center}
\end{figure}
\subsection{Perturbations}
%
We initially perturb
the above equilibrium impulsively by a Gaussian pulse in the component of velocity that is nearly parallel to ambient magnetic field lines, $V_{\rm ||}$,
viz.,
\beq\label{eq:perturb}
V_{\rm ||}(x,y,t=0) = A_{\rm v} \exp\left[ -\frac{(x-x_{\rm 0})^2+(y-y_{\rm 0})^2}{w^2} \right]\, .
\eeq
Here $A_{\rm v}$ is the amplitude of the pulse, $(x_{\rm 0},y_{\rm 0})$ is its initial position and
$w$ denotes its width. We set and hold fixed $A_{\rm v}=40$ km s$^{-1}$, $w=150$ km, and $y_{\rm 0}=1.75$ Mm, while allow $x_{\rm 0}$ to vary.
We consider two cases: (a) $x_{\rm 0}=0$ Mm and (b) $x_{\rm 0}=5$ Mm.
\section{Results of numerical simulations}
Equations (\ref{eq:MHD_rho})-(\ref{eq:MHD_B}) are solved numerically using the code FLASH
(Lee \& Deane 2009). This code implements a second-order unsplit Godunov solver (e.g., Murawski \cite{mur02})
with various slope
limiters and Riemann solvers, as well as adaptive mesh refinement (AMR).
We use the minmod slope limiter and the Roe Riemann solver (e.g., Toro 2009).
For the case of (a) and (b)
we set the simulation box as 
$(5,5)\, {\rm Mm} \times (1,41)\, {\rm Mm}$ and $(0,30)\, {\rm Mm} \times (1,31)\, {\rm Mm}$, 
respectively.
We impose boundary conditions by fixing in time all plasma quantities
at
all four boundaries
to there equilibrium values.
In all our studies we use 
a static but non-uniform
grid with a minimum (maximum) level of
refinement set to $1$ ($7$). See Fig.~\ref{fig:amr} for the grid system for the case of $x_{\rm 0}=5$ Mm. A similar grid system was chosen
for the case of $x_{\rm 0}=0$ Mm.
As the grid was static no
refinement strategy
was adopted.
As each block consists of $8\times 8$ identical numerical cells, we reach 
the effective spatial resolution of about $60$ km. 
We discuss two cases: (a) essentially vertical magnetic field and (b) oblique magnetic field. For the case (a)
we launch the initial pulse at $x_{\rm 0}=0$ Mm, while the case (b) is realized for $x_{\rm 0}=5$ Mm. In the latter case magnetic field
creates the angle of about $\pi/3$ to the horizontal direction.
%
%
\subsection{Essentially vertical magnetic field}
The temporal snapshots of the simulated macrospicule in the
model atmosphere of the VAL-C temperature and essentially vertical magnetic field are shown in Figure~\ref{fig:spicule_prof_cent}.
It displays the spatial profiles of plasma temperature (colour maps) and velocity (arrows),
resulting from the initial velocity pulse that was launched at $x_{\rm 0}=0$ Mm.
The upper left panel corresponds to $t=100$ s.
Cold chromospheric plasma lags behind the shock front,
which at this time reaches altitude of $y\simeq 8$ Mm. 
The reason for the material being lifted up is 
the rarefaction of the plasma behind the shock front, 
which leads to low pressure there (Hansteen et al. \cite{hansten2006}). 
As a result, the pressure gradient works against gravity and forces
the chromospheric material to penetrate the solar corona.

It is noteworthy that small-amplitude vertically propagating waves are described by the Klein-Gordon equation (Lamb \cite{lamb}).
It follows from this equation that initial, spatially-localized perturbation results in a wavefront which moves away from the launching place.
The disturbance ahead of the wavefront is at rest, but behind the wavefront an oscillation wake results in. This wake oscillates
with essentially constant frequency and the amplitude of the oscillations behind the wavefront declines gradually in time. As a consequence of the equilibrium mass density fall off with height, the amplitude of these oscillations
grows with altitude. Then, the linear theory is not valid anymore as nonlinear effects result in trailing shocks with the secondary shock
following the leading shock (Hollweg \cite{hol82}).

The next snapshot (top middle panel) is drawn for $t=200$ s, when the macrospicule raised to the altitude of $y\simeq 11$ Mm.
At $t=300$ s (top-right panel)
the macrospicule already subsided.
As the secondary shock lifts up
the chromospheric material (see two small peaks on both sides of the macrospicule),
there are upward flows at the macrospicule sides, well seen at $t=400$ s (middle-left panel).
It is clear from the snapshots at $t\ge 200$ s
that the plasma
exhibits some horizontal oscillations.
This is because the plasma pushes the magnetic field lines outside due to the large value of plasma $\beta$ at the chromosphere.
The signatures of these horizontal oscillations are
the surface waves which are running along transition region-corona interface.
There are well seen bi-directional flows in the macrospicule: downward at the centre and upward at the boundaries,
which agrees with the observational data
(Tsiropoula et al. \cite{Tsiropoula1994}, Tziotziou et al. \cite{Tziotziou2003,Tziotziou2004}, Pasachoff et al. \cite{pas09}).
Eventually, in case of our numerical simulation, the downward streaming
plasma is also following a path in the cool core region
as seen at $t=800$ s. 
Since the macrospicule plasma at its leading edge enters in the corona, it may be heated quickly at
the coronal temperature, and associated with the greater upflow velocity typically in the
range of $20-40$ km s$^{-1}$ compared to the background in case of both singular and bi-directional
flows (cf., Fig. 3). Therefore, this segment of macrospicule may
appear in form of Doppler blue-shift of the centroid of FUV/EUV lines formed
with maximum ionic fraction at their typical coronal/TR formation temperatures.
The blue shift and plasma upflows with the typical macrospicule rise-up speed have been observed
by Wilhelm (\cite {klaus2000}) and Suematsu et al. (\cite{Suematsu1995}).
However, the scenario of temperature, flow, and magnetic field
structures in the bottom part of the macrospicules are rather complex.
The cool core, steep temperature gradient towards boundary may evolve
at the base. The up and down-flow structuring (cf. Fig.~\ref{fig:spicule_prof_cent})
may be observed in form of mixed red and blue-shift scenario
as previously explained by Wilhelm (\cite {klaus2000})
in the pass-band of cool chromospheric lines.

The middle-right panel is drawn
for $t=600$ s, which clearly resembles a macrospicule-like structure with
the chromospheric temperature and mass density. Its width and mean rising speed are about $0.4$ Mm and
$30$ km s$^{-1}$, respectively. These values are close to the corresponding characteristics of typical macrospicules.
However, this is just a presentation of the simulation case study of the
solar macrospicule in essentially vertical magnetic field configuration. The typical length, width, speed, and life time can vary
by tuning the initial conditions of the numerical model.
Afterwards the material again falls back at the central part, while the next shock forces the plasma to move upwards
at the spicule boundaries.
We see quasi-recurrently accuring single ($t=200$ s, $t=600$ s, $t=800$ s), double (e.g., $t=400$ s), and triple (e.g. $t=300$ s) structures
which are characteristic features of macrospicules.

Figure~\ref{fig:time_profile} illustrates the
vertical
component
of velocity that is collected in time at the detection point $(x=0, y=14)$ Mm for
the case of Fig.~\ref{fig:spicule_prof_cent}.
As a result of fast mass density fall off with height upwardly propagating waves
grow in their amplitude and
steepen rapidly into shocks.
The arrival of the first shock front at the detection point, $y=14$ Mm, is clearly seen at $t\simeq 70$ s.
The second shock front reaches the detection point at $t\simeq 460$ s, i.e., after $\sim 290$ s.
This secondary shock results from the nonlinear wake, which lags behind the leading shock.
\subsection{Oblique magnetic field}
The temporal snapshots of the simulated macrospicule in the
model atmosphere of the VAL-C temperature and oblique magnetic field are shown in
Figure~\ref{fig:spicule_prof_x0=5}. It illustrates the spatial profiles of plasma temperature
(colour maps) and vertical velocity (arrows)
resulting from the initial velocity pulse that was launched at $x_{\rm 0}=5$ Mm.
As a result, the plasma flows along the inclined magnetic field.
The dynamics of plasma essentially resembles that obtained by vertically propagating pulse.
At $t=200$ s the
velocity signal reaches the altitude of
up to {\bf $8$} Mm.
Then it falls back again resembling the consecutive double structure and bi-directional flows.

Figure~\ref{fig:time_profile_x0=5} illustrates the vertical
component
of velocity that is collected in time at the detection point $(x_d=8.5, y_d=14)$ Mm for
the case of Fig.~\ref{fig:spicule_prof_x0=5}.
The arrival of the first shock front at the detection point
is clearly seen at $t\simeq 100$ s.
The second
shock arrives to
the detection point at $t=430$ s i.e., after $\sim$ 330 s. 
Another two shocks appear at latter times that is at $\sim 650$ s and $\sim900$ s. 
%

%

However, this is also a presentation of the simulation case study of the
solar macrospicule in oblique magnetic field configuration driven by a velocity
pulse in the model temperature of the solar atmosphere.
The typical length, width, speed, and life time of the
obliquely propagating macrospicules can also vary
by tuning the initial conditions of the numerical model.
\section{Observational evidence of a pulse driven macrospicule}
\begin{figure*}
\centering
\begin{tabular}{ccc}
\includegraphics[scale=0.50, angle=90]{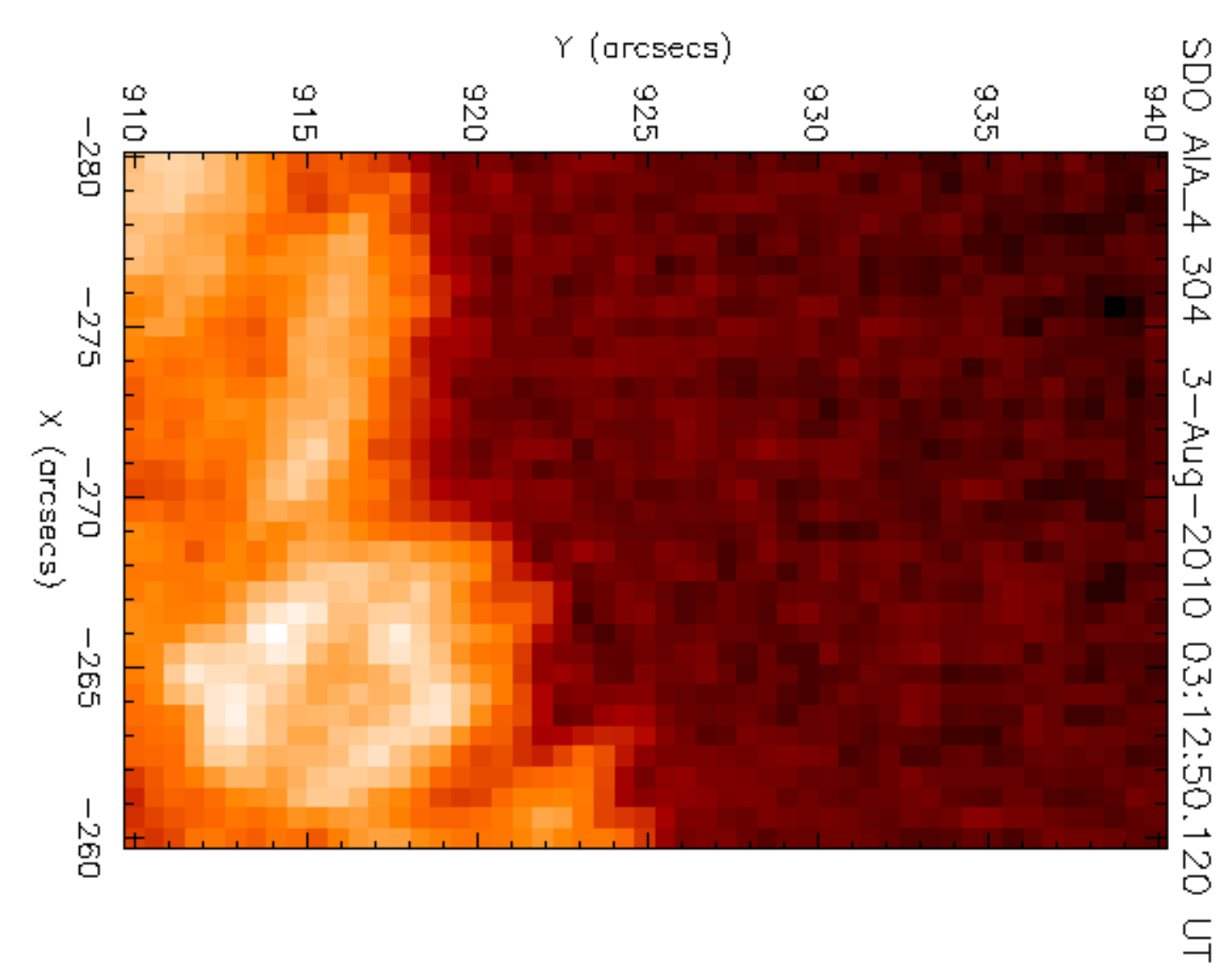} &
\includegraphics[scale=0.50, angle=90]{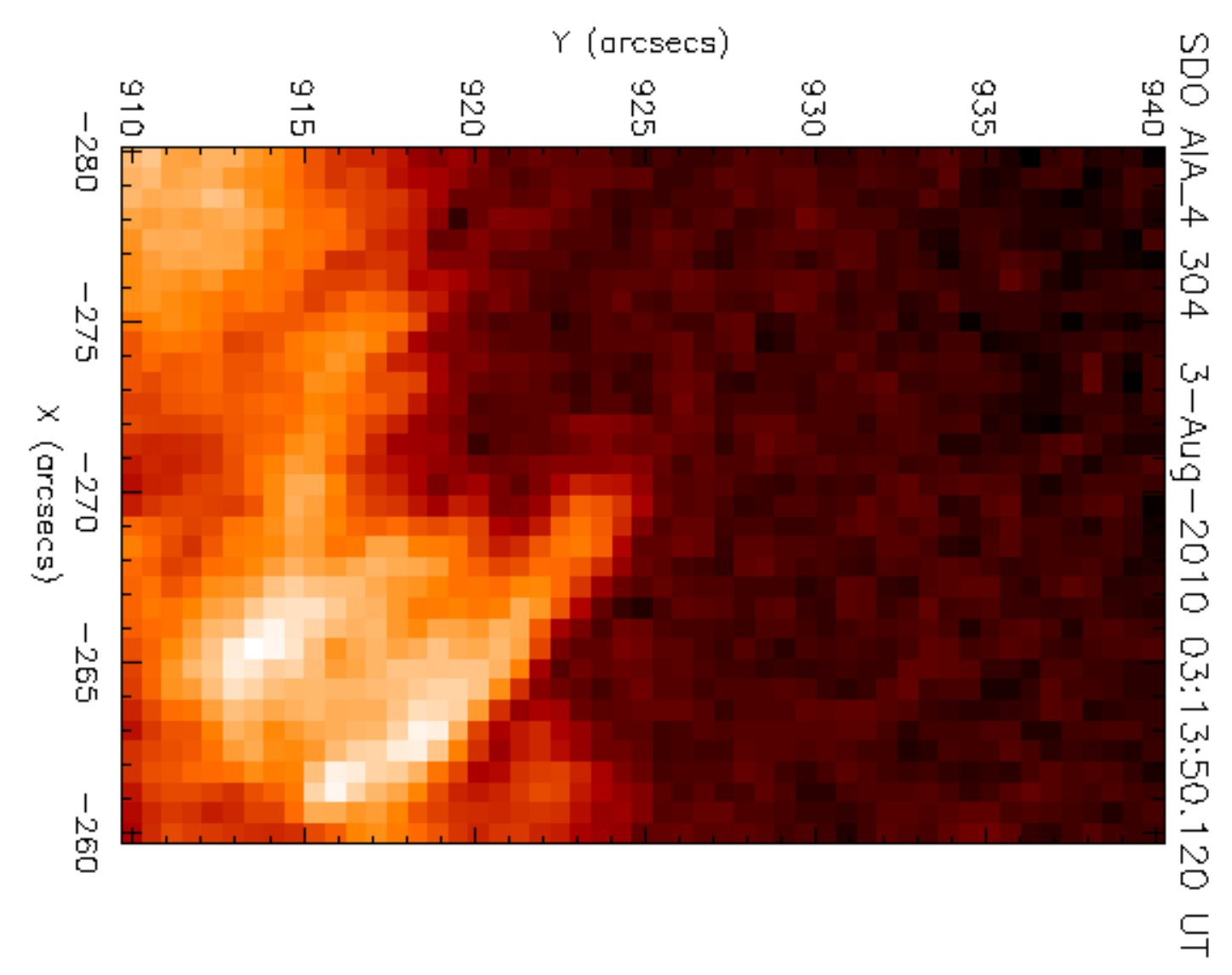} &
\includegraphics[scale=0.50, angle=90]{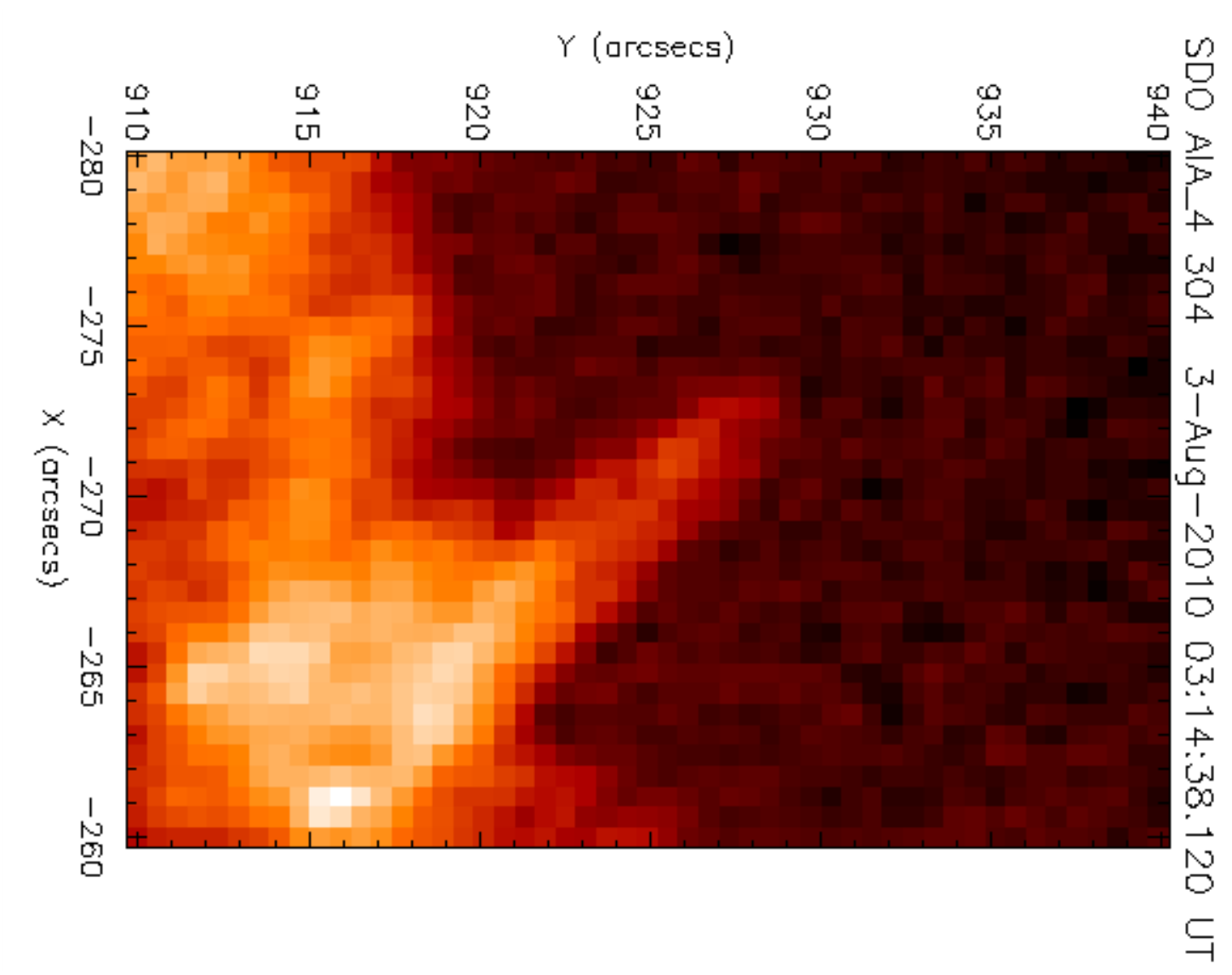} \\
\includegraphics[scale=0.50, angle=90]{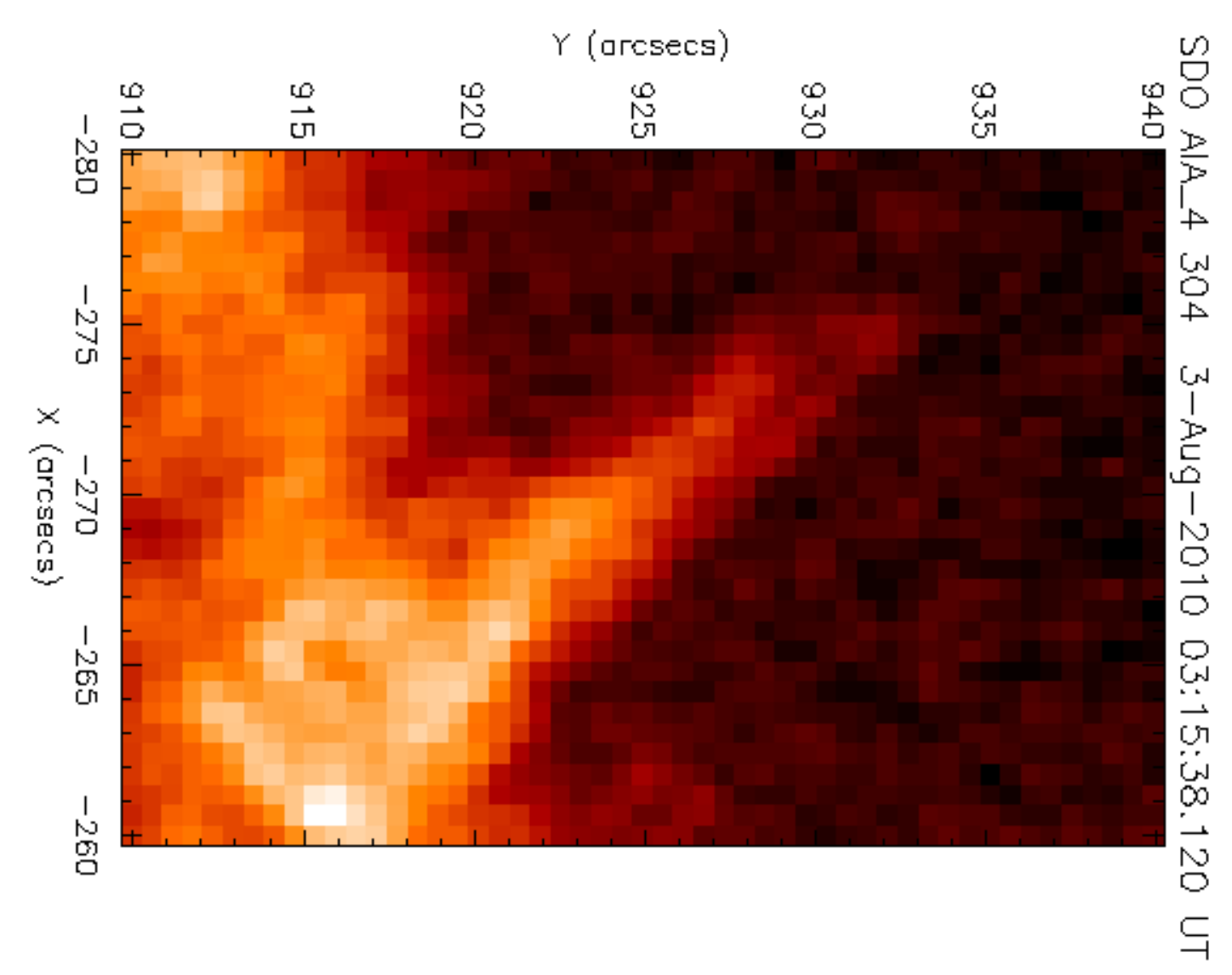} &
\includegraphics[scale=0.50, angle=90]{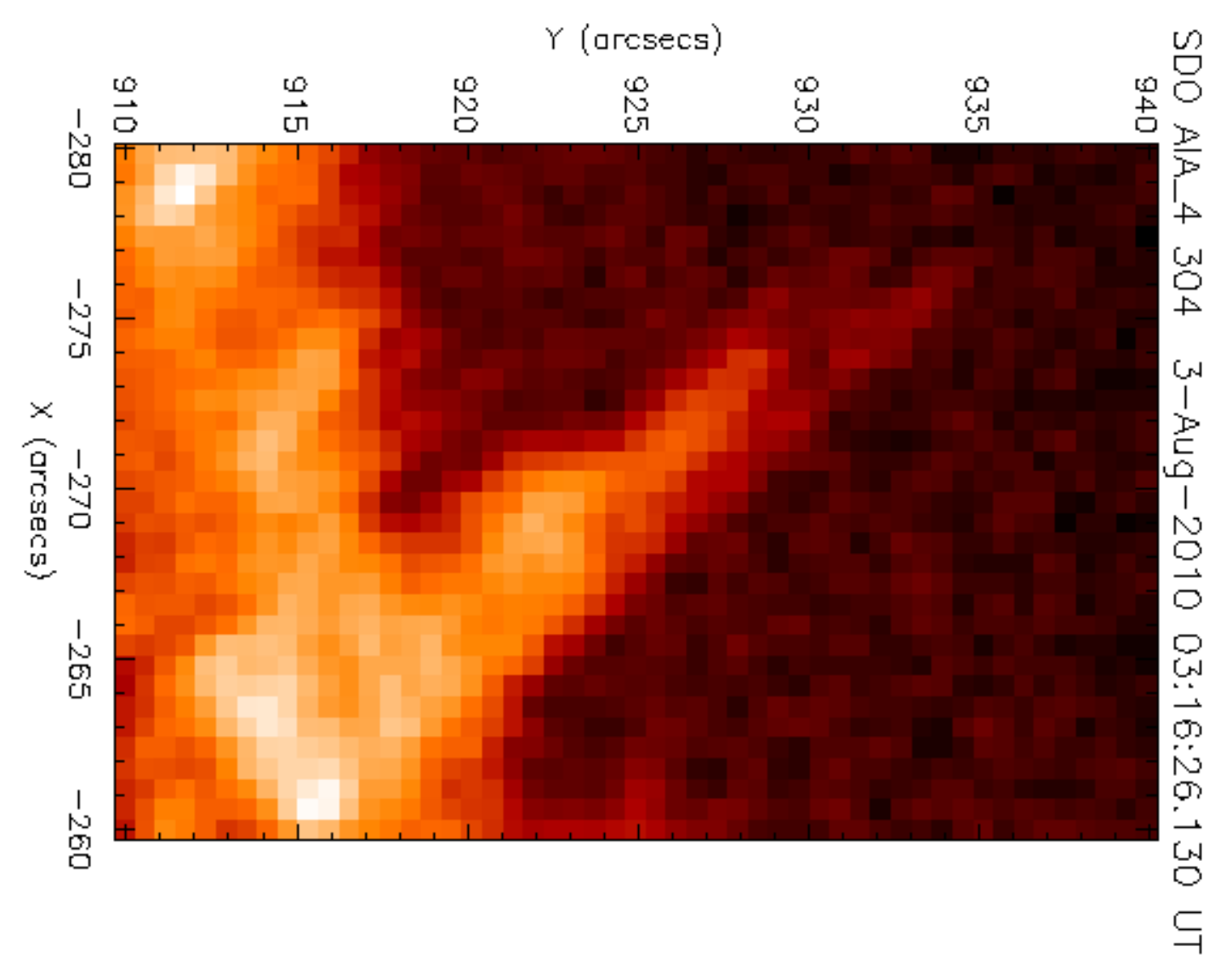} &
\includegraphics[scale=0.50, angle=90]{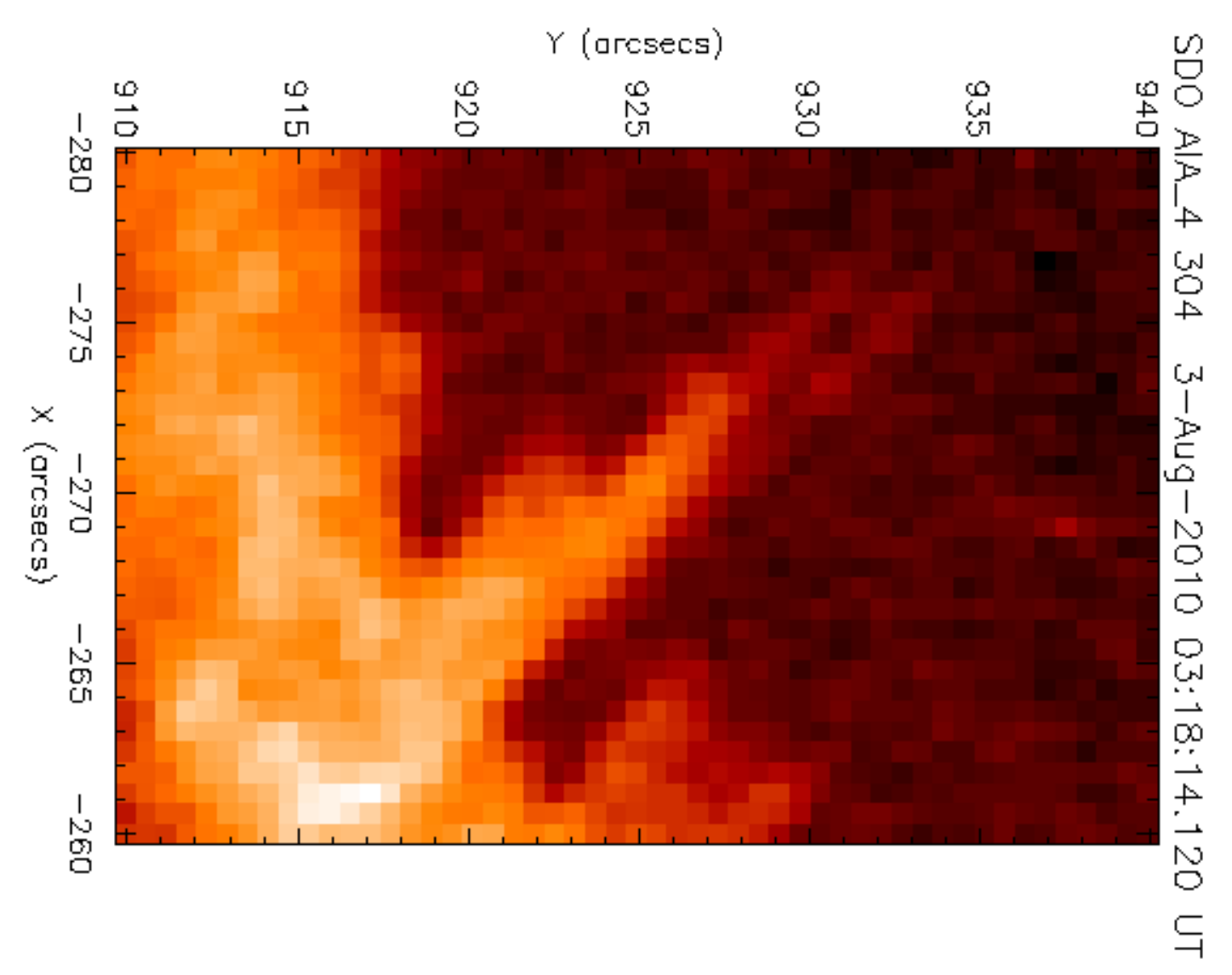} \\
\includegraphics[scale=0.50, angle=90]{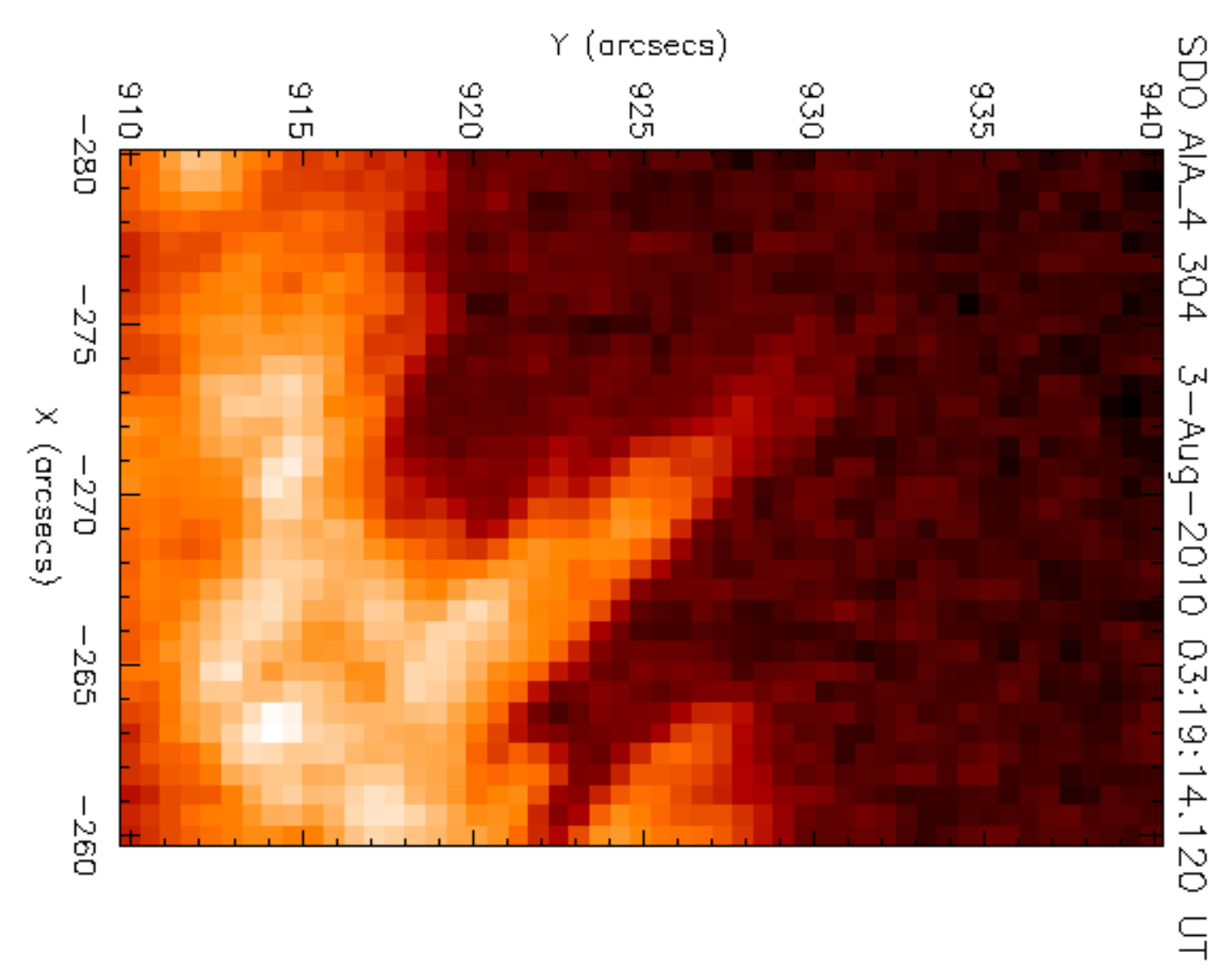} &
\includegraphics[scale=0.50, angle=90]{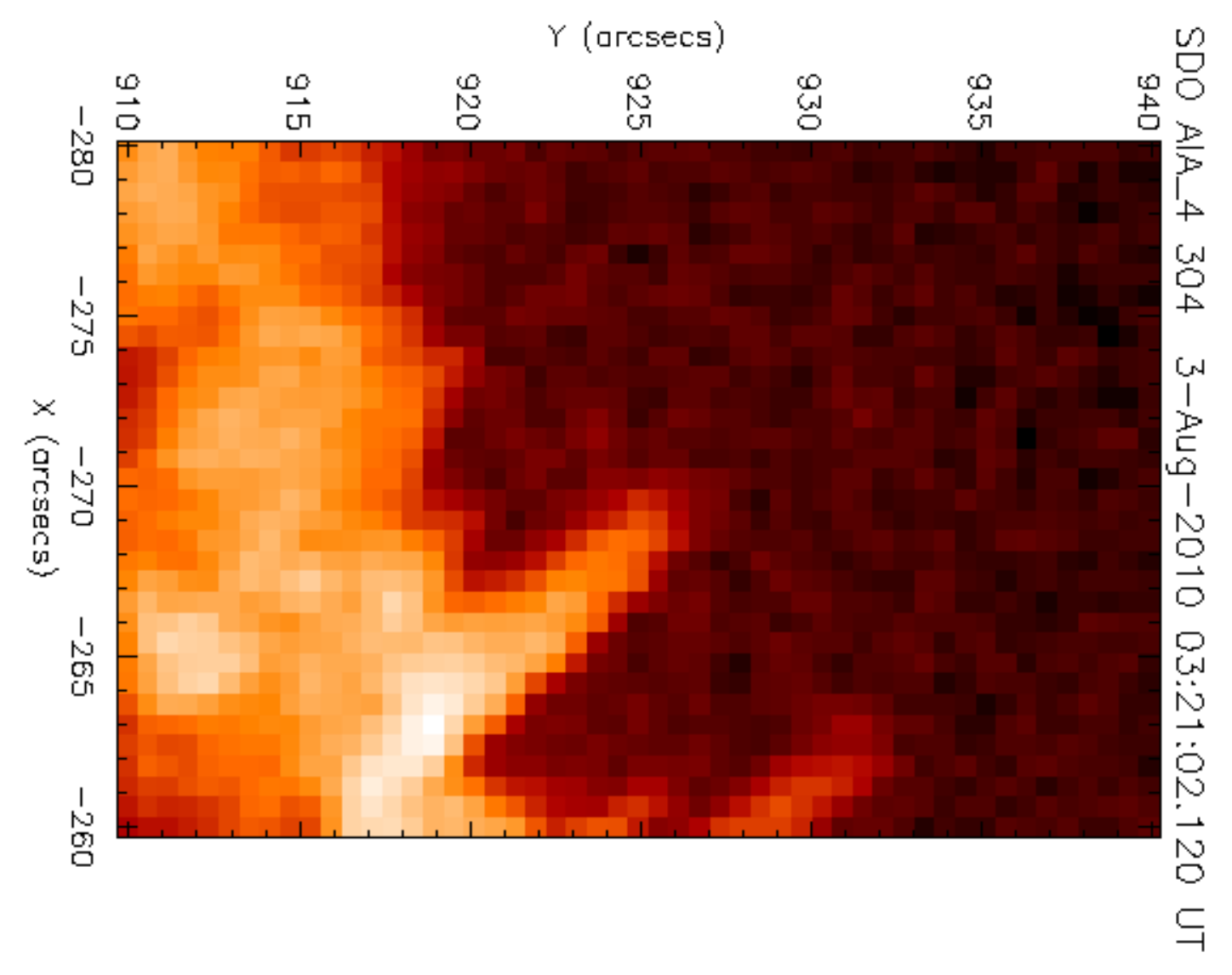} &
\includegraphics[scale=0.50, angle=90]{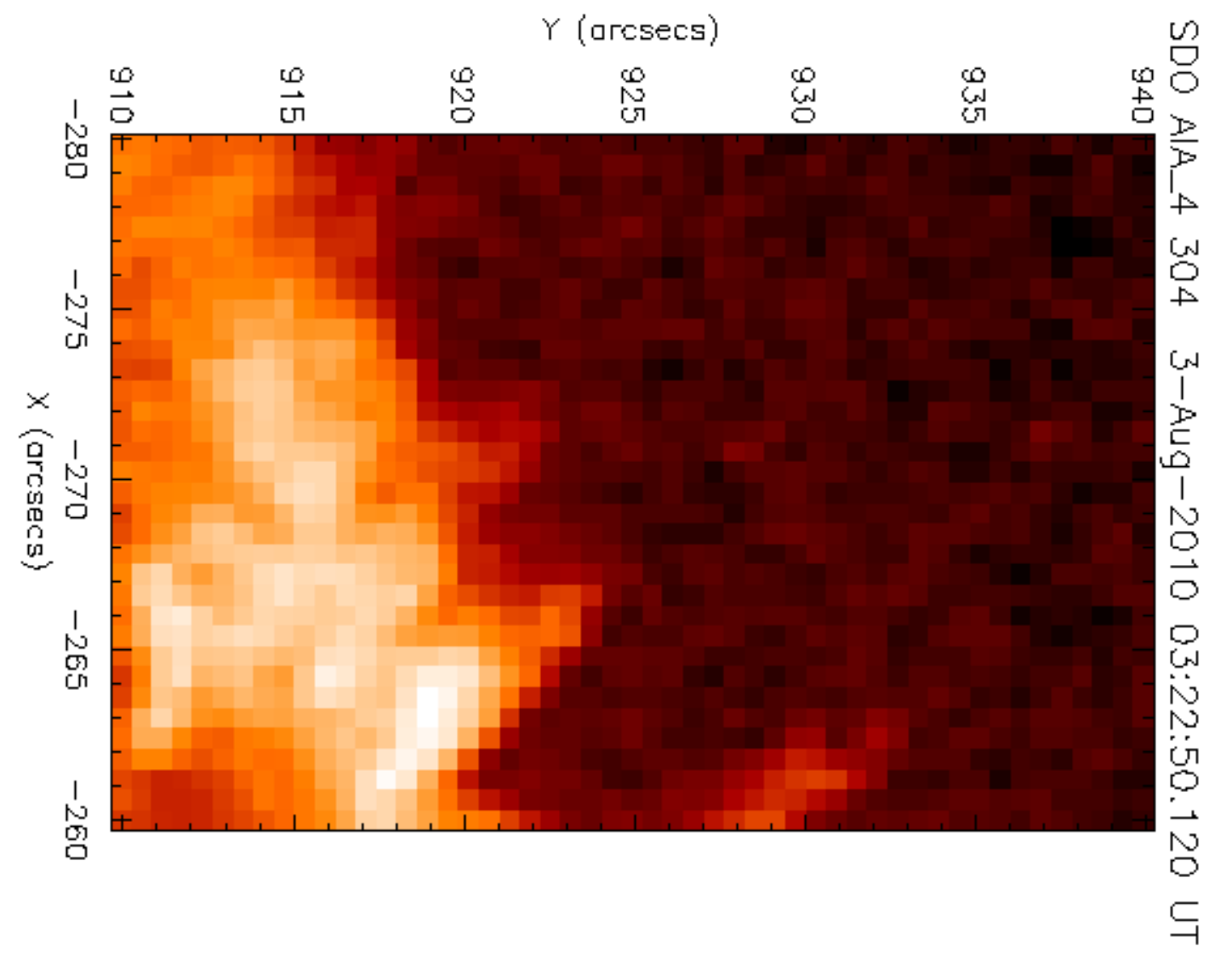} \\
\end{tabular}
\caption{\small
The SDO/AIA 304 \AA\ time sequence of a macrospicule propagation above the solar limb in the north-east polar coronal hole that
clearly shows the upward motion of the plasma and its downward motion along the same path.
}
\label{fig:JET-PULSE}
\end{figure*}
In the present section, we describe an observational signature of a limb spicule which
approximately resembles the simulated macrospicule as driven by a velocity pulse in the
more general oblique magnetic field configuration.
We use a time-series data of a solar spicule at north polar coronal hole
as observed in 304 \AA \ filter of Atmospheric
Assembly Imager (AIA) onboard the Solar Dynamics Observatory (SDO)
on 3 August 2010 during 03:12:50 UT--03:22:50 UT.
The SDO/AIA has a typical resolution of 0.6$"$ per pixel
and the highest cadence of 12 s, and it observes the
full solar disk in three UV (1600 \AA, 1700 \AA, 4500 \AA)
and seven EUV (171 \AA, 193 \AA, 211 \AA, 94 \AA, 304 \AA, 335 \AA
, 131 \AA) wavelengths. Therefore, it provides
the unique observations of multi-temperature, high-resolution,
and high-temporal plasma dynamics all over the Sun.
The field-of-view of the north polar coronal hole as observed
by SDO/AIA 304 \AA\ on 03 August 2010 was (1094$"$, 317$"$), while
the (X$_{cen}$, Y$_{cen}$)
was (-29.618$"$, 918.631$"$).
The time series has been obtained by the SSW cutout service at LMSAL, USA, which
is corrected for the flat-field and spikes. We run aia\_prep subroutine
of SSW IDL also for further calibration and cleaning of the time
series data.

We observe a giant spicule dynamics at the limb of north
polar coronal hole.
The macrospicule has been launched obliquely
from the background open field lines of the polar coronal
hole.
The life-time of the spicule was observed as $\sim$10 min, which fits with the
typical life time of the spicules/macrospicules (Georgakilas et al. \cite{georg99}, Sterling \cite{Sterling2000}, Pasachoff et al. \cite{pas09}).
The observed macrospicule reaches upto a height of $\sim$12 Mm and has the width of
$\sim$2 Mm.
It should be noted that the present observation is only a case study in the support
of the numerical simulation of a solar macrospicule. Macrospicules
may reach to 7-40 Mm heights with a typical life time of 3-45 mins
(Sterling \cite{Sterling2000}).
Our aim here is only to show that these observations reveal
upto some extent the
nature of the spicule as per the numerical modeling.

The spicule starts moving up
above the solar limb on $\sim$03:13 UT and reaches at a maximum height
of $\sim$12 Mm at $\sim$03:17 UT with approximate average rising speed of $\sim$45 km s$^{-1}$.
The height of $\sim$12 Mm, life-time of
$\sim$10 min, and the speed of $\sim$45 km s$^{-1}$ are typical for the
observed macrospicules (Georgakilas et al. \cite{georg99}, Sterling \cite{Sterling2000}).
The unique evidence this macrospicule presented is its rise obliquely
off the limb during $\sim$250 s of its life time, and then its
fall back along the same path. 
The simulated macrospicule in the oblique magnetic field configuration (cf., Fig.~5) also reaches at 
a height of $\sim$8 Mm in $\sim$200 s, and then its material is attempting to fall back.
The falling material of the simulated macrospicule is encountered
by the wave train of the another upcoming pulses that uplift
the spicule material quasi-periodically upto a certain
height again and again in the solar atmosphere.
The observed dynamics of macrospicule also supports that it may be formed by a similar velocity
pulse that triggered in the chromosphere and steepens in form of a shock
in the transition region/corona. 
It should be noted that the rising time of the observed
macrospicule is $\sim$250 s, however, its falling time
is larger as $\sim$350 s. This means that the falling cool
plasma may be interacting with the uplifted plasma
coming from the lower solar atmosphere along the same path
of the spicule. It is also clear from the
snapshots during 03:16 UT-03:23 UT that the falling plasma
is trying to settle down, while some brightened material is being pushed up
from the lower part of atmosphere.
However, the signal in the line 304 \AA \ filter is not necessarily optically thick, and other
contributions from the line-of-sight signal may contaminate the observations.
Therefore, it is not completely clear that the brightening at the footprint is due to the rebound compression.
After reaching in the upper atmosphere along the magnetic field lines,
this spicule falls back
and cool plasma
traced back its backward path as observed here.

Despite of these uncertainties,
the observed feature is approximately similar to
the simulated macrospicule in the oblique magnetic field.
In the case of the observed spicule, finally the plasma is being
settled down. The reason may be that the upcoming pulses in the real atmosphere
may not have the same amplitude and spatio-temporal distribution
at the base of the spicule. The decay of the wave train
of the pulses does not probably allow the quasi-periodic
rise and fall of the spicule plasma from the same place.
The downflowing material may also
suppress the upcoming pulses that may
finally allow the settling down of the spicule plasma.

The observed spicule is two times wider than the simulated
one. However, the mean rising time and maximal height are
almost closer for observed and simulated macrospicules. 
The height of the simulated spicule
can be tuned exactly to the observations by small adjustment of
the velocity pulse. Simulated and observed velocities are
also matching to the typical velocities of the macrospicules.
The fine tuning of the initial conditions
in the numerical simulation can exactly mimic the observed conditions,
however, here our aim is only to present
an observational case study of the macrospicule to reveal
its pulse driven nature.
The resolution and real complex conditions of the
solar atmosphere do not permit us exactly to precisely study the quasi-periodic
rise and fall of the spicule plasma at the same place in the observations as
were evident in the simulations. However, greater life time of the
falling plasma, and some observational evidence in the time series
(03:17--03:23 UT) indicate interaction of falling material
with the element of rising plasma. This may be due to the arrival of another
pulse from below the chromosphere, which may try to push the
plasma opposing the gravitational free  fall.

In conclusions, the observations of the macrospicule
resembles the simulation of pulse driven macrospicules.
One more interesting point is noticeable in the observations that
the spicule is being generated near the boundary of a brightened
network at north polar coronal hole. This network may be the place
of the launching of a velocity pulse that triggers
the macrospicule higher in the solar atmosphere. It should be noted that
all the spicules/macrospicules do not show the similar properties as
we observe here. Therefore, the activity in the associated brightened
magnetic network may be the most probable cause to drive the initial
velocity pulse to trigger the observed macrospicule.
However, this is just a case study to support the
numerical simulations presented in the paper, and the measured spatial and temporal
scales are approximate when compared to the simulations.
The detailed observational search of the
solar macrospicules matching exactly with the numerical
simulations, will be done in our future projects.
\section{Discussion and conclusions}
The formation of solar macrospicules is still an unresolved problem in solar physics. Several competitive mechanisms have been supposed from time to time,
but none of them could explain all properties of macrospicules. We performed 2D numerical simulation of the velocity pulse,
which was launched at the chromosphere, in stratified solar atmosphere with the VAL-C temperature profile.
The amplitude of upward propagating
perturbation
rapidly grows with height due to the rapid decrease of
equilibrium mass
density.
Therefore, the perturbation quickly steepens into the shock in upper regions of the solar chromosphere that launches the cool
spicule material behind it (Hansteen et al. \cite{hansten2006}). 
The simulated properties of the macrospicules both in the vertical and oblique magnetic field
configurations, e.g., velocity, height, life-time, bi-directional flows etc, clearly match
with the typical properties of such type of giant spicules.

We also present a case study of the observed macrospicule using the recent observations
from SDO/AIA in 304 \AA\ to
support the observations.
The observed macrospicule rises obliquely off the limb in north polar coronal hole and reaches
at a height of $\sim$12 Mm in the first 3-4 minutes of its total life time of $\sim$10 min.
These observed phenomena and parameters
support the simulated macrospicule
in the oblique magnetic field configuration. The width of the observed spicule is, however,
comparatively larger that the simulated one. But, this
is one of the free parameters of the numerical simulation that can be
tuned with the shape and size of the velocity pulse to trigger such macrospicules.
The amplitude of the velocity pulse can also tune the rising speed of the simulated spicule.
The observed spicule rises up and then falls back along the same path
just like the simulated pulse driven macrospicule. The doubly splitted head
of the observed spicule is also evident in the observations
once it reaches near the maximum height. This may be
the
indication of the initiation of bi-directional plasma flows
along the spicule as similar to the numerical simulations.
Although, the baseline of the observations does not allow
us to examine the interaction of the falling spicule material
with the uplifting plasma. However, the longer falling phase compared to the rising one
may indicate some interaction between falling and rising plasmas.
Unfortunately, we do not have the Doppler velocity
fields and related observations to comment more on this aspect. However,
we can easily identify some upward motion of dense bright plasma
near the spicule base around 03:16-03:21 UT that may oppose the
falling plasma that is already being lifted upto a maximum
height of $\sim$12 Mm. When such interaction occurs, the falling plasma
is probably being subsided and shows the two splitted heads as
a double component as also reported in the simulation
snapshot of 400 s in Fig. 5.

In conclusions, our numerical simulations
of the solar macrospicules
exhibit
approximately observed properties.
We should admit that
the performed 2D simulations are ideal in the sense that
they do not include radiative transfer, thermal conduction along
field lines etc. The field
configuration and the initial stratification are simple and in
hydrostatic equilibrium. Since we did not attempt to
synthesize observables, the exact comparison with observations is
unlikely.
The detailed observational search of the
solar macrospicules matching exactly with the numerical
simulations, will be done in our future projects.
Simultaneous
imaging and spectroscopic observations
should also be performed in future to explore
extensively the dynamics of such macrospicules that will
impose rigid constraints on the numerical simulation
models also. 

\begin{acknowledgements}
The authors express their thanks to the referee for his/her comment. 
KM thanks Kamil Murawski for his assistance in drawing numerical data. 
AKS acknowledges Shobhna Srivastava for the patient encouragements. 
The work of TZ was supported by the Austrian Fonds zur F\"orderung der
wissenschaftlichen Forschung (project P21197-N16) and the Georgian National Science
Foundation (under grant GNSF/ST09/4-310).
The software used in this work was in part developed by the DOE-supported ASC/Alliance Center for
Astrophysical Thermonuclear Flashes at the University of Chicago.

\end{acknowledgements}


%

\end{document}